\documentclass[preprint,amsmath]{revtex4-1}

\usepackage[utf8]{inputenc}
\usepackage{dutchcal}
\usepackage{amsmath}
\usepackage{amssymb}
\usepackage{color}
\usepackage[normalem]{ulem}
\usepackage{mathtools}
\usepackage{color}
\usepackage{graphicx}
\usepackage{subcaption}

\begin{document}
\title{ The Transient Case of The Quenched Trap Model}
\author{Stanislav Burov}
\email{stasbur@gmail.com}
\affiliation{Physics Department, Bar-Ilan University, Ramat Gan 5290002,
Israel}

\pacs{PACS}

\begin{abstract}
 In this work the diffusion in the quenched trap model with diverging mean waiting times is examined. The approach of randomly stopped time is extensively applied in order to obtain asymptotically exact representation of the disorder averaged positional probability density function. 
 We establish that the dimensionality and the geometric properties of the lattice, on top of which the disorder is imposed, dictate the plausibility of a mean-filed approximation that will only include  annealed disorder. 
 Specifically, for any case when the probability to return to the origin ($Q_0$) is less than $1$, i.e. the transient case, the quenched trap model can be mapped on to the continuous time random walk. The explicit form of the mapping is provided. 
 In the case when an external force is applied on a tracer particle in a media described by the quenched trap model, the response to such force is calculated and a non-linear response for sufficiently low dimensionality is observed.
 \end{abstract}

\maketitle

\section{Introduction}

Brownian Motion is probably the simplest manifestation of a transport in random environment. In this case the particle path is constantly modified by collisions with molecules that compose the surrounding media. The trajectory will appear as if the direction of motion is randomly changes as a function of time and a simple random walk (RW) is quite useful to describe the motion. 
 The continuum representation of a RW is a regular diffusion~\cite{Weiss}. When the motion of the particle occurs in a complex media, the simple RW might be insufficient for proper description of the transport. In many materials the basic linear dependence of the mean squared displacement (MSD), $\langle x^2(t) \rangle$, is missing and instead $\langle x^2(t)\rangle\sim t^{\alpha} $ while $0<\alpha<1$. Such behavior is termed anomalous subdiffusion and materials where it appears include living cells~\cite{Metzler2011,LiveCell,Tabei,Bariers}, blinking quantum dots~\cite{QuantumD}, plasma membrane~\cite{Krapf2011}, filamentous networks~\cite{BurovPnas} and many more~\cite{Sokolov2005}. 
The modeling of transport in these systems is quite complicated, when compared to the original RW. In the works of Scher and Montroll~\cite{ScherMontroll} the continuous time random walk (CTRW) approach for transport in amorphous materials was developed. The idea behind CTRW is the existence of regions of local arrest, i.e. traps, where the traced particle waits for some random time before it continues its motion inside the media. 
When the expected random waiting times diverge the behavior is non-ergodic~\cite{Bel2005,YongHe} and CTRW will produce the mentioned subdiffusive scaling of the MSD.   
While CTRW became extremely popular and applicative~\cite{Bouchaud,Klafter,Kutner2017}, this approach treats the disorder in the media as annealed and uncorrelated. 
Quenchness of the disorder in the media is more physically appealing in many situations but it implies existence of strong correlations that in their turn introduce significant difficulties in calculating basic properties of the transport~\cite{Kehr}. When the local dwell times of CTRW are fixed the model is known as the quenched trap model (QTM).

The QTM was found to be an important model that describes glassy behavior such as aging, weak-ergodicity breaking and non self-averaging~\cite{BouchaudAg,Monthus1996,Rinn2000,Rinn2001,Bertin,Burov2007}. 
Beyond the applications of the QTM, the difficulty of untangling the behavior dictated by quenched disorder, that is associated with QTM, posed this model and methods for its solution as a fundamental problem of anomalous transport~\cite{Bouchaud}. 
The presence of the mentioned correlations, imposed by the quenched disorder, make the treatment of the QTM highly non-trivial task. 
Over the years many theoretical methods were devised to advance the general understanding of the QTM. The method of semi-equilibration~\cite{Derrida} allowed to determine the average velocity and diffusion constant in  the one-dimensional ($d=1$) case for the non-anomalous transport. 
Description of the QTM in terms of master equation and their representation in the Fourier space produced the scaling behavior of the QTM propagator at the origin~\cite{Bernasconi, Alexander}. 
Renormalization Group approach~\cite{Machta}, and scaling arguments~\cite{Bouchaud1987}, provided the existence of a critical dimension, $d=2$, for the QTM and the scaling behavior of the MSD. 
Based on these works a qualitative understanding that for sufficient high dimension ($d>2$) the behavior of the QTM can be mapped on-to the mean filed representation, i.e. CTRW. 
Further, the behavior of the QTM was studied for various lattices under the simplification of directed walk, i.e. without returns to previously visited traps~\cite{Aslangul}. 
The decimation of disorder allowed Monthus to calculate (among other quantities) the behavior of the positional probability density function (PDF) in $d=1$ case in the limit of very low temperatures~\cite{Monthus,MonthusSec}. Rigorous probabilistic approach to the QTM led to mathematically exact scaling theorems~\cite{BenArous1,BenArous2} and further generalization of the QTM to such models as the randomly trapped random walk~\cite{BenArous3,Cerny01}. The effect of fractal structures for QTM~\cite{Akimoto2015} and behavior of the QTM under influence of a bias~\cite{Akimoto2019} are part of a current research.

The previously obtained results suggest that for any dimension $d>2$ the behavior of QTM converges to the one of CTRW. A simple hand-waving argument that support this qualitative result is that in sufficiently high dimensions the traced particle rarely returns to the same lattice point, thus reducing the effect of strong correlations imposed by the quenched disorder. The P{\'o}lya's~\cite{Weiss} theorem states that the probability to return to the origin (or any previously occupied position) is less then $1$ for any dimension above $d=2$. 
A valid question is what is the quantitative representation of the mapping between QTM and CTRW? Can one extend this mapping to the cases where dimensionality is low but the formerly raised hand-waiving argument still holds, i.e. the biased case?
In this manuscript we will provide an explicit form of the mapping between QTM and CTRW for any transient case in any dimension.  By using the randomly stopped time approach, that was originally  developed for the $d=1$ case~\cite{Burov1,Burov2}, we manage to obtain a subordiantion of the spatial process to the temporal $\alpha$-stable process. Unlike the CTRW where the subordinated spatial process advances as a function of the number of jumps~\cite{Bouchaud,Fogedby,Barkai}, for QTM the local time of the spatial process is quite different. A brief summary of part of our results was published in Ref.~\cite{Burov2017}.

This paper is organized as follows. In Sec.~\ref{section_def} the QTM is defined together with local time, measurement time and the subordination approach. In Sec.~\ref{salphaSec} the local time $S_\alpha$ is explored and the mean value of the local time is computed in Sec.~\ref{meansalpha} and the second moment in Sec.~\ref{secondsalpha}. In Sec.~\ref{deltafunction} we summarize the results of the first and second moment calculation and show that the local time convergences to the number of jumps that the process has performed. In Section~\ref{doublesubordination} the previously established convergence of the local time is exploited in order to establish an explicit mapping between the CTRW and QTM, by the means of double subordination. The formulas are applied to the one-dimensional cased of biased QTM. In Sec.~\ref{nonlinresp} we obtain analytic expressions for the moments of the transient case of the QTM and show how the quenched  disorder gives rise to the non-linear response of externally applied field. The summary is provided in Sec.~\ref{summary}. Several Appendices supply specific technical calculations and referred to in the manuscript.  

\section{The Quenched Trap Model and Subordination}
\label{section_def}

The QTM is defined as a random jump process of a particle on top of a lattice of dimension $d$. For every lattice point ${\bf x}$ a quenched random variable $\tau_{\bf x}$ is defined. 
This quenched variable $\tau_{\bf x}$ defines the time that the particle is going to spend at ${\bf x}$ before jumping to some other site ${\bf x}'$', i.e. $\tau_{\bf x}$ is the local dwell time. The probability to jump from ${\bf x}$ to ${\bf x}'$ is provided by $p({\bf x}',{\bf x})$. In the following we will assume translational invariance of the lattice that leads to $p({\bf x}',{\bf x})$ of the form $p({\bf x}'-{\bf x})$.  The quenched dwell times $\{\tau_{\bf x}\}$ are , real, positive and independently distributed random variables with 
\begin{equation}
    \psi(\tau_{\bf x})\sim\tau^{-(1+\alpha)}A\big/|\Gamma(-\alpha)|\qquad \left(\tau_{\bf x}\to\infty\right)
    \label{psitaudef}
\end{equation}
as the PDF ($A>0$). The value of the exponent $\alpha$ is bounded to $0<\alpha<1$. For such values of $\alpha$ the average dwell time is diverging, $\int_0^\infty\tau\psi(\tau)\,d\tau\to\infty$ and the model gives rise to anomalous subdiffusion and aging~\cite{BouchaudAg}. 
The physical picture behind this definition of QTM is a thermally activated particle that is jumping between various energetic traps. When a particle is in a trap, the average escape time $\tau$ is provided by the Arrhenius law $\tau\propto \exp\left(E_{\bf x}/T\right)$, where $E_{\bf x}>0$ is the depth of the trap ${\bf x}$ and $T$ is the temperature. When the distribution of $E_{\bf x}$s is   $f(E)=\frac{1}{T_g}\exp\left(-E{\bf x}/T_g\right)$, the average escape time is distributed according to Eq.~(\ref{psitaudef}), and $\alpha=T/T_g$. For low temperatures $T<T_g$ and glassy behavior, i.e. aging and non-self-averaging, is observed~\cite{Bertin}. The QTM is thus a version of a transport on top of a random energetic landscape with exponential distribution of trap depths.

We wish to perform a separation of the QTM into two processes. The first one is a spatial process on top of the lattice. This process is defined by the jump probabilities $p({\bf x}'-{\bf x})$ with some local time. The other process is a temporal process that transforms the local time into  the measurement time $t$, that is defined by the dwell times. How exactly the measurement time and the local time are defined and related to each other is crucial for the solution of the QTM.

\subsection{Measurement Time and Local Time}
\label{loctime}

During the measurement time $t$, the particle has visited several lattice points and stayed exactly $\tau_{\bf x}$ at each lattice ${\bf x}$. 
The measurement time $t$ is then simply given by 
\begin{equation}
t=\sum_{\bf x} n_{\bf x}\tau_{\bf x}
    \label{measurtime}
\end{equation}
where $n_{\bf x}$ is the number of time the particle visited site ${\bf x}$ and the summation is over all the lattice points. While $\tau_{\bf x}$ are independent, identically distributed (I.I.D) random variables, $n_{\bf x}$ are correlated. Indeed, the number of times the particle visited at site ${\bf x}$ shouldn't be very different from the number of times the particle visited in  adjacent sites. The local time for the spatial process is defined as 
\begin{equation}
S_\alpha=\sum_{\bf x} \left(n_{\bf x}\right)^\alpha.
    \label{localtime}
\end{equation}
The variable 
\begin{equation}
\eta=t/(S_\alpha)^{1/\alpha}
    \label{etadef}
\end{equation}
 is of high interest, especially in the $t\to\infty$ and $S_\alpha\to\infty$ limit. Lets consider that $\{n_{\bf x}\}$ are fixed (an outcome of a given experiment) then $\eta$ depends on the realization of the disorder, i.e. $\{\tau_{\bf x}\}$. The PDF of $\eta$ is found by examining disorder averaged $\exp(-u\eta)$, i.e $\langle \exp\left(-u\eta\right)\rangle$, that is given by
\begin{equation}
    \langle e^{-u\eta} \rangle =\displaystyle \langle \exp\left( -u \sum_{\bf x}\frac{n_{\bf x}\tau_{\bf x}}{(S_\alpha)^{1/\alpha}}\right) \rangle.
    \label{etalaplace}
\end{equation}
Since the $\{\tau_{\bf x}\}$ are I.I.D Eq.~(\ref{etalaplace}) takes the form
\begin{equation}
\langle e^{-u\eta} \rangle =\displaystyle \prod_{\bf x} {\hat{\psi}}\left[\frac{n_{\bf x}u}{(S_\alpha)^{1/\alpha}}\right],
    \label{etalaplace02}
\end{equation}
where the product is over all the lattice sites and ${\hat{\psi}}(u)=\int_0^\infty\exp(-\tau_{\bf x} u)\psi(\tau_{\bf x})\,d\tau_{\bf x}$.
Due to Eq.~(\ref{psitaudef}) the small $u\to 0$ limit of ${\hat{\psi}}(u)$ is ${\hat{\psi}}(u)\sim 1-Au^\alpha$ and Eq.~(\ref{etalaplace02}) takes the form
\begin{equation}
\langle e^{-u\eta} \rangle =\displaystyle \prod_{\bf x} \left( 1-\frac{n_{\bf x}^\alpha}{S_\alpha}Au^\alpha\right).
    \label{etalpalace03}
\end{equation}
When all the multiplications are performed on the r.h.s. of Eq.~(\ref{etalpalace03}) the leading term is $1$. The next term is $-\sum_{\bf x}n_{\bf x}^\alpha Au^\alpha/S_\alpha$ that is simply $-A u^\alpha$. The following term is $\frac{1}{2}\sum_{\bf x}\sum_{{\bf x}'}n_{\bf x}^\alpha n_{{\bf x}'}^\alpha A^2 u^{2\alpha}/S_\alpha^2$ that takes the form $\frac{1}{2}A^2u^{2\alpha}$. By computing  next terms with higher orders of $u$ we obtain that the r.h.s is of the form $\sum_{j=0}^\infty(-Au^\alpha)^j/\Gamma[j+1]$, that is simply the Taylor  expansion of $\exp\left(-Au^\alpha\right)$. When taking into account the higher orders of $u$ in the expansion of ${\hat{\psi}}(u)=1-Au^\alpha+Bu^\beta+...$ (where $\beta>\alpha$), we show in  Appendix~\ref{sbetaproof} that in the limit of $S_\alpha\to\infty$ all these terms converge to $0$ and do not contribute to the r.h.s of Eq.~(\ref{etalpalace03}). Finally we can state that in the large $S_\alpha$ limit 
\begin{equation}
\langle e^{-u\eta} \rangle = e^{-A u^\alpha}
    \label{etalaplacefnl}
\end{equation}
which means that the PDF of $\eta$ is one sided L{\'e}vy stable distribution $l_{\alpha,A,1}$~\cite{Klafter,Barkai}. We managed to obtain the distribution of $\eta$ and the distribution of the measurement time $t$ for a given local time $S_\alpha$, since  $t=S_\alpha^{1/\alpha}\eta$. Because $S_\alpha$ is positive and strictly growing, as we let the particle jump from one lattice point to another, we can inverse the relation in Eq.~(\ref{etadef}), $S_\alpha=(t/\eta)^\alpha$, use the known distribution of $\eta$, and obtain the PDF of $S_\alpha$ for a given measurement time $t$
\begin{equation}
    {\cal P}_t\left(S_\alpha\right)\sim
    \frac{t}{\alpha}S_\alpha^{-1/\alpha-1}l_{\alpha,A,1}\left(\frac{t}{S_\alpha^{1/\alpha}}\right)
    \label{salphadist}
\end{equation}
in the large $t$ limit. The measurement time $t$ is the quantity that is set in any experiment or calculation. Eq.~(\ref{salphadist}) describes the probability to obtain various $S_\alpha$ when averaging over disorder and letting the process to evolve up to time $t$. We use this disorder-averaged relation between local time $S_\alpha$ and $t$ in the next subsection while constructing the representation of the QTM propagator in terms of  the two processes. 

\subsection{Subordination}
\label{Subordintaion}

The probability $p({\bf x}'-{\bf x})$ describes the transition probability between two lattice points. It completely determines the spatial process on top of any translationally-invariant lattice, as long as we don't take the disorder due to traps into account. 
For example it determines the probability to find the particle at position ${\bf x}$ after $N$ jumps. In this case, $N$ is the local time of the spatial process, the process is terminated when the number of performed jumps reaches a specific threshold and the position is recorded. 
Any strictly growing function of the jumps can be considered as a local time, specifically $S_\alpha$. 
When the process starts $S_\alpha$ equals to zero and its value is updated each time the particle performs a jump. As  $S_\alpha$ crosses a given value the process is terminated. The quantity $P_{S_\alpha}({\bf x})$ is the probability to find the particle at position ${\bf x}$ (starting from the origin) after local time $S_\alpha$ has passed. 
Due to dependence of $S_\alpha$ on local visitation numbers $n_{\bf x}$ (Eq.~(\ref{localtime})), the local time is a function of both the number of jumps and the trajectory taken by the particle. 

The PDF $P({\bf x},t)$ to find the particle at position ${\bf x}$ after measurement time $t$ is presented by conditioning on all the possible $S_\alpha$ that can occur during the process. 
One needs to sum over all the possible $P_{S_\alpha}({\bf x})$ multiplied by the appropriate probability to observe such $S_\alpha$ at time $t$, for a given disorder. After averaging over disorder the PDF takes the form 
\begin{equation}
\langle P({\bf x},t) \rangle=\sum_{S_\alpha}P_{S_\alpha}({\bf x}) 
{\cal P}_t(S_\alpha)
    \label{subordination01}
\end{equation}
and due to Eq.~(\ref{salphadist}) in the $t\to\infty$ limit we obtain
\begin{equation}
\langle P({\bf x},t) \rangle\sim\int_0^\infty P_{S\alpha}({\bf x})
\frac{t}{\alpha}S_\alpha^{-1/\alpha-1}l_{\alpha,A,1}\left(\frac{t}{S_\alpha^{1/\alpha}}\right)\,dS_\alpha.
    \label{subordination02}
\end{equation}
while we replaced the summation by integral~\cite{Bouchaud}. Eq.~(\ref{subordination02}) represents the propagator of the QTM as a subordination of two processes. The spatial process that has no disorder but is terminated at random local time $S_\alpha$ and the temporal process that involves the disorder and make the mapping between local time and measurement time. While the function $l_{\alpha,A,1}(\dots)$ is known, the missing part is the probability $P_{S_\alpha}$ that is obtained for the case of a transient spatial process.

\section{Local time $S_\alpha$}
\label{salphaSec}

The propagator $P_{S_\alpha}({\bf x})$ lacks the disorder that is present in the QTM and is a simple jump process on a lattice, but nevertheless it is highly non-trivial. 
The main complication is the stopping time $S_\alpha$ that is dependent on the path taken by the particle.
If the local time is simply the number of jumps $N$, the probability to find the particle at ${\bf x}$ after $N$ jumps is completely defined by the corresponding probabilities after $(N-1)$ jumps. This is not the case for $P_{S_\alpha}({\bf x})$. 
The arrival to ${\bf x}$ do not increases $S_\alpha$ by $1$ as happens with the number of jumps, but rather the increase of $S_\alpha$ depends on the total number of times that ${\bf x}$ was previously visited. 
In the case of $1$-dimensional simple random walk (RW) the shape of $P_{S_\alpha}({\bf x})$ was previously~\cite{Burov1} computed in the limit of $\alpha\to 0$. In this example $P_{S_\alpha}({\bf x})$ has a very distinctive V shape (with a minimum at the origin) and is quite different from the regular Gaussian propagator of the random walk.

Before obtaining the $P_{S_\alpha}({\bf x})$, the study of the properties of $S_\alpha$ is in place. Specifically the first two moments of $S_\alpha$, i.e. ${\overline{S_\alpha}}$ and ${\overline{S_\alpha^2}}$. The averaging ${\overline{\dots}}$ is with respects to many trajectories of the RW walk on a lattice without traps. The results of Sec.\ref{meansalpha} and Sec.\ref{secondsalpha} are summarized in Sec.\ref{deltafunction}.

\subsection{${\overline{S_\alpha}}(N)$}
\label{meansalpha}

The mean value of $S_\alpha$ is obtained from Eq.~(\ref{localtime}), ${\overline{S_\alpha}}=\sum_{\bf x}\overline{n_{\bf x}^\alpha}$. Defining $\beta_N({\bf x};k)$ to be the probability for the RW to visit lattice site ${\bf x}$ exactly $k$ times after $N$ steps, we write the average local time after $N$ steps as
\begin{equation}
{\overline{S_\alpha}}(N)=\sum_{\bf x}\sum_{k=0}^\infty k^\alpha \beta_N({\bf x};k).
    \label{salphamean01}
\end{equation}
The probability $\beta_N({\bf x};k)$ is the probability to arrive at ${\bf x}$ at-least $k$ times minus the probability to arrive at-least $k+1$ times during $N$ jumps. Since the $k$th arrival must occur during these $N$ jumps,  $\beta_N({\bf x};k)$ is expressed as
\begin{equation}
    \begin{array}{ll}
      \beta_N({\bf x};k)=\sum_{m=1}^N f_m({\bf x};k) - \sum_{m=1}^N f_m({\bf x};k+1) 
      & \qquad {\bf x}\neq {\bf 0} 
      \\
      \beta_N({\bf 0};k)=\sum_{m=1}^N f_m({\bf 0};k-1) - \sum_{m=1}^N f_m({\bf 0};k) 
      & 
    \end{array}
    \label{betaxk01}
\end{equation}
where $f_N({\bf x};k)$ is the probability to reach site ${\bf x}$ for the $k$'th time after $N$ steps. By defining $f_N({\bf 0})$ to be the probability of first return to the origin (${\bf x}={\bf 0}$) after $N$ steps, we write the recursive form for $f_N({\bf x};k)$
\begin{equation}
f_N({\bf x};k+1)=\sum_{m=0}^N f_m({\bf x};k)f_{N-m}({\bf 0}).
\label{firstpassagesdef}
\end{equation}
The generating function ${\hat f}_z({\bf x};k)=\sum_{N=0}^\infty z^N f_N({\bf x};k)$ is then
\begin{equation}
{\hat f}_z({\bf x};k)=\left[{\hat f}_z({\bf 0})\right]^{k-1}{\hat f}_z({\bf x})
    \label{frecursive}
\end{equation}
where ${\hat f}_z({\bf 0})$ is the generating function of the probability of first return to ${\bf 0}$ and ${\hat f}_z({\bf x})$ is the generating function of the probability of first arrival to ${\bf x}$. Eq.~(\ref{betaxk01}) and Eq.~(\ref{frecursive}) provide the generating function of $\beta_N({\bf x};k)$
\begin{equation}
    \begin{array}{ll}
      {\hat \beta}_z({\bf x};k)=\frac{1}{1-z}\left[1-{\hat f}_z({\bf 0})\right]\left[{\hat f}_z({\bf 0})\right]^{k-1}{\hat f}_z({\bf x}) 
      & \qquad {\bf x}\neq {\bf 0} 
      \\
      {\hat \beta}_z({\bf 0};k)=\frac{1}{1-z}\left[1-{\hat f}_z({\bf 0})\right]\left[{\hat f}_z({\bf 0})\right]^{k-1} 
      & 
    \end{array}
    \label{betaxk02}
\end{equation}
Eq.~(\ref{betaxk02}) allows us to compute the generating function of ${\overline{S_\alpha}}(N)$, while the summation $\sum_{\bf x}{\hat f}_x({\bf x})$ can be obtained by the means of $c_N({\bf x})$, the probability to find the particle at position ${\bf x}$ after $N$ steps (started at ${\bf 0}$). Since $c_N({\bf x})$ is related to $f_N({\bf x})$ by 
\begin{equation}
c_N({\bf x}) = \delta_{N,0}\delta_{{\bf x},{\bf 0}} + 
\sum_{m=1}^N f_m({\bf x}) c_{N-m}({\bf 0})
    \label{cnxdefinition}
\end{equation}
the generating functions ${\hat f}_z({\bf x})$ and ${\hat c}_z({\bf x})$ are connected by 
\begin{equation}
    \begin{array}{l}
    {\hat f}_z({\bf x\neq 0}) = {\hat c}_z({\bf \neq 0})\big/{\hat c}_z({\bf 0}) 
    \\
    {\hat f}_z({\bf 0}) =1- 1\big/{\hat c}_z({\bf 0}) .
    \end{array}
    \label{candfgenerating}
    \end{equation}
Together with the fact that $\sum_{\bf x} c_N({\bf x}) =1$ and consequently $\sum_{\bf x} {\hat c}_z({\bf x})=1/(1-z)$, Eqs.(\ref{salphamean01},\ref{betaxk02},\ref{candfgenerating}) result in 
\begin{equation}
{\overline {{\hat{S}_\alpha}}}(z) = \left[\frac{1-{\hat f}_z({\bf 0})}{1-z} \right]^2 \sum_{k=0}^\infty k^\alpha {\hat f}_z({\bf 0})^{k-1}.
\label{salphameanz01}
\end{equation}
For the case when the spatial process is transient and the probability of eventually returning to the origin $Q_0=\sum_{N=0}^\infty f_N({\bf 0})$, is less than $1$, the asymptotic ($N\to\infty$) is readily obtained from Eq.~(\ref{salphameanz01}). For $z\to 1$, ${\hat f}_z({\bf 0})\to Q_0<1$. The fact that $\sum_{N=0}^\infty N z^N = z/(1-z)^2$ and Tauberian theorem~\cite{Weiss} implies that
\begin{equation}
    {\overline {S_\alpha}}(N)\sim \Lambda N \qquad (N\to\infty)
    \label{salphaNlarge}
\end{equation}
where 
\begin{equation}
    \Lambda = \frac{\left[1-Q_0\right]^2}{Q_0} Li_{-\alpha}(Q_0)
    \label{lambdaconst}
\end{equation}
and $Li_a(b)=\sum_{k=1}^\infty b^k/k^a$ is the Polylogarithm function.

The form of average $S_\alpha$ as expressed in Eq.~(\ref{salphaNlarge}) will be essential in the following for asymptotic representation of $\langle P({\bf x},t) \rangle$ for the transient case by the means of $P_{S_\alpha}({\bf x})$. The average behavior of $S_\alpha$ suggests that the local time $S_\alpha$ is not very much different from the regular local time, i.e. the number of jumps $N$, at-least for the transient case $Q_0<1$. The behavior of the second moment of $S_\alpha$ should indicate if one indeed can exchange the local time $S_\alpha$ by a linear function of $N$.

\subsection{${\overline{S_\alpha^2}}(N)$}
\label{secondsalpha}

The goal of this section is to provide the conditions for a plausible substitution of $S_\alpha$ by its average value $\langle S_\alpha\rangle$. The second moment of $S_\alpha$ is computed in a similar fashion as the first moment was computed in Sec.~\ref{meansalpha}, and the first moment of $S_\alpha$ (Eq.~\ref{salphamean01}) is generalized to 
\begin{equation}
{\overline{S_\alpha^2}}(N) = \sum_{{\bf x}}\sum_{{\bf x}'}\sum_{k_1}\sum_{k_2}k_1^\alpha k_2
^\alpha \beta_N\left({\bf x};k_1,{\bf x}';k_2\right)
\label{secmomdef}
\end{equation}
where $\beta_N\left({\bf x};k_1,{\bf x}';k_2\right)$ is the probability that in $N$ steps the RW will visit site ${\bf x}$ exactly $k_1$ times and the site ${\bf x}'$ exactly $k_2$ times. This probability is calculated in the terms of $f_N\left({\bf x},k_1;{\bf x}',k_2\right)$, the probability to arrive to ${\bf x}$ after $N$ steps for the $k_1$th time while visiting ${\bf x}'$ exactly $k_2$ times. $\beta_N\left({\bf x};k_1,{\bf x}';k_2\right)$ is the probability that the $k_1$th arrival was performed but not the $(k_1+1)$th, i.e.
\begin{equation}
\begin{array}{ll}
     \beta_N\left({\bf x};k_1,{\bf x}';k_2\right)= &
\sum_{l=0}^N \left\{\left[
f_l\left({\bf x},k_1;{\bf x}',k_2\right)-f_l\left({\bf x},k_1+1;{\bf x}',k_2\right)
\right]\right.
       \\
       &
    \left. +\left[
f_l\left({\bf x}',k_2;{\bf x},k_1\right)-f_l\left({\bf x}',k_2+1;{\bf x},k_1\right)
\right]\right\}\qquad (k_1+k_2\geq 2).
\end{array}
\label{twopointbeta01}
\end{equation}
The range $k_1>0$ and $k_2>0$ is sufficient since $\beta_N({\bf x},k_1;{\bf x'},k_2)$ is multiplied by $k^\alpha$ in Eq.~(\ref{secmomdef}). We define the probability to start at ${\bf x}$ and after $N$ steps to reach ${\bf x}'$, without visiting ${\bf x}$ or ${\bf x}'$ on the way, as $M_N({\bf x},{\bf x}')$ and the probability to start at ${\bf x}$ and return to the same site after $N$ steps, without visiting ${\bf x}$ or ${\bf x}'$ on the way, as $T_N({\bf x},{\bf x}')$. 
The probability $f_N({\bf x},k_1;{\bf x}';k_2)$ is recursively expressed in terms of $M_N({\bf x},{\bf x}')$ and $T_N({\bf x},{\bf x}')$
\begin{equation}
\begin{array}{ll}
f_N({\bf x},k_1+1;{\bf x}';k_2)= &
\sum_{l=0}^N 
f_l({\bf x},k_1;{\bf x}';k_2)T_{N-l}({\bf x},{\bf x}')+f_l({\bf x}',k_2;{\bf x};k_1)M_{n-l}({\bf x}',{\bf x})
\\
\end{array}
    \label{frstpsgmt01}
\end{equation}
where $f_N({\bf x},0;{\bf x'},k_2)=0$. Eq.~(\ref{frstpsgmt01}) leads to the following expression in $z$ space 
\begin{equation}
{\hat f}_z({\bf x},k_1+1;{\bf x}',k_2)=
{\hat f}_z({\bf x},k_1;{\bf x}',k_2){\hat T}_z({\bf x},{\bf x}')+
{\hat f}_z({\bf x}',k_2;{\bf x},k_1){\hat M}_z({\bf x}',{\bf x}).
    \label{frstpsgmt02}
\end{equation}
Application of additional transformation $k_1\to\xi_1$ and $k_2\to\xi_2$, by performing a double summation $\sum_{k_1=1}^\infty\sum_{k_2=1}^\infty\xi_1^{k_1}\xi_2^{k_2}$ on both sides of Eq.~(\ref{frstpsgmt02}) delivers
\begin{equation}
\left[
1-\xi_1{\hat T}_z({\bf x},{\bf x'})
\right]
{\hat {\tilde f}}_z({\bf x},\xi_1;{\bf x}',\xi_2) 
-
\xi_1{\hat M}_z({\bf x}',{\bf x})
{\hat {\tilde f}}_z({\bf x}',\xi_2;{\bf x},\xi_1)
=\xi_1{\hat f'}_z({\bf x},1;{\bf x'},\xi_2)
    \label{frstpsgxi01}
\end{equation}
where ${\hat {\tilde f}}_z({\bf x},\xi_1;{\bf x}',\xi_2)=\sum_{k_1=1}^\infty\sum_{k_2=1}^\infty\xi_1^{k_1}\xi_2^{k_2}{\hat f}_z({\bf x},k_1;{\bf x}',k_2)$ and \\ ${\hat f'}_z({\bf x},1;{\bf x'},\xi_2)=\sum_{k_2=1}^\infty\xi_2^{k_2}{\hat f}_z({\bf x},1;{\bf x}',k_2)$.
In a similar fashion we obtain 
\begin{equation}
\left[
1-\xi_2{\hat T}_z({\bf x'},{\bf x})
\right]
{\hat {\tilde f}}_z({\bf x'},\xi_2;{\bf x},\xi_1) 
-
\xi_2{\hat M}_z({\bf x},{\bf x'})
{\hat {\tilde f}}_z({\bf x},\xi_1;{\bf x'},\xi_2)
=\xi_2{\hat f'}_z({\bf x'},1;{\bf x},\xi_1).
    \label{frstpsgxi02}
\end{equation}
Eqs.~(\ref{frstpsgxi01},\ref{frstpsgxi02}) are linear equations in terms of ${\hat{\tilde f}}_z({\bf x},\xi_1;{\bf x'},\xi_2)$ and ${\hat{\tilde f}}_z({\bf x'},\xi_2;{\bf x},\xi_1)$ that attain the solution 
\begin{equation}
      {\hat{\tilde f}}_z({\bf x},\xi_1;{\bf x'},\xi_2) = 
      \frac{\xi_1\left[1-\xi_2 {\hat T}_z({\bf x'},{\bf x})\right]{\hat f'}_z({\bf x},1;{\bf x'},\xi_2)+\xi_1\xi_2{\hat M}_z({\bf x'},{\bf x}){\hat f'}_z({\bf x'},1;{\bf x},\xi_1)}
      {[1-\xi_1{\hat T}_z({\bf x},{\bf x'})]
      [1-\xi_2{\hat T}_z({\bf x'},{\bf x})]
      -\xi_1\xi_2{\hat M}_z({\bf x},{\bf x'}){\hat M}_z({\bf x'},{\bf x})}.
    \label{genformf01}
\end{equation}
Since $f_N({\bf x'},k+1;{\bf x},0)=\sum_{l=0}^N f_l({\bf x'},k;{\bf x},0) T_{N-l}({\bf x'},{\bf x})$, the transform ${\hat{ {f'}}}_z({\bf x'},\xi_2;{\bf x},0)=\sum_{k=1}^\infty \xi_2^k {\hat f}_z({\bf x'},k;{\bf x},0)$ is 
\begin{equation}
    {\hat{\tilde {f'}}}_z({\bf x'},\xi_2;{\bf x},0) =
    \frac{\xi_2 {\hat f}_z({\bf x'},1;{\bf x},0)}{1-\xi_2 {\hat T}_z({\bf x'},{\bf x})}.
    \label{specformf02}
\end{equation}
By using the expression $f_N({\bf x},1;{\bf x'},k_2)=\sum_{l=0}^N f_l({\bf x'},k_2;{\bf x},0) M_{N-l}({\bf x'},{\bf x})$ and Eq.~(\ref{specformf02}) we obtain 
\begin{equation}
    {\hat{\tilde{f'}}}_z({\bf x},1;{\bf x'},\xi_2) = 
     \frac{\xi_2 {\hat f}_z({\bf x'},1;{\bf x},0)
     {\hat M}_z({\bf x'},{\bf x})}
     {1-\xi_2 {\hat T}_z({\bf x'},{\bf x})},
    \label{specformf03}
\end{equation}
and then by substitution of Eqs.~(\ref{genformf01},\ref{specformf03}) in Eq.~(\ref{twopointbeta01}), and using Eq.~(\ref{frstpsgmt02}), we obtain for ${\hat{\tilde{\beta}}}_z({\bf x};\xi_1,{\bf x'};\xi_2)=\sum_{N=0}^\infty\sum_{k_1=1}^\infty\sum_{k_2=1}^\infty z^N {\xi_1}^{k_1}{\xi_2}^{k_2} \beta_N({\bf x};k_1,{\bf x'};k_2)$
\begin{equation}
    \begin{array}{ll}
         {\hat{\tilde{\beta}}}_z({\bf x};\xi_1,{\bf x'};\xi_2)=
         & 
         \frac{1}{1-z}\left\{
         \left(
         1-{\hat T}_z({\bf x},{\bf x'})-{\hat M}_z({\bf x'},{\bf x})
         \right)
         \frac{\xi_1\xi_2 {\hat f}_z({\bf x'},1;{\bf x},0)
     {\hat M}_z({\bf x'},{\bf x})+{\xi_1}^2\xi_2{\hat M}_z({\bf x'},{\bf x})\frac{ {\hat f}_z({\bf x},1;{\bf x},0)
     {\hat M}_z({\bf x},{\bf x'})}
     {1-\xi_1 {\hat T}_z({\bf x},{\bf x'})}}
      {[1-\xi_1{\hat T}_z({\bf x},{\bf x'})]
      [1-\xi_2{\hat T}_z({\bf x'},{\bf x})]
      -\xi_1\xi_2{\hat M}_z({\bf x},{\bf x'}){\hat M}_z({\bf x'},{\bf x})}\right.
         \\
         &
         \left.+
       \left(
         1-{\hat T}_z({\bf x'},{\bf x})-{\hat M}_z({\bf x},{\bf x'})
         \right)
         \frac{\xi_2\xi_1 {\hat f}_z({\bf x},1;{\bf x'},0)
     {\hat M}_z({\bf x},{\bf x'})+{\xi_2}^2\xi_1{\hat M}_z({\bf x},{\bf x'})\frac{ {\hat f}_z({\bf x'},1;{\bf x'},0)
     {\hat M}_z({\bf x'},{\bf x})}
     {1-\xi_2 {\hat T}_z({\bf x'},{\bf x})}}
      {[1-\xi_1{\hat T}_z({\bf x'},{\bf x})]
      [1-\xi_1{\hat T}_z({\bf x},{\bf x'})]
      -\xi_2\xi_1{\hat M}_z({\bf x'},{\bf x}){\hat M}_z({\bf x},{\bf x'})}
      \right\}
       .
    \end{array}
    \label{twopointbeta02}
\end{equation}
The generating functions of the two-point probabilities $T_N({\bf x},{\bf x'})$, $M_N({\bf x},{\bf x'})$ and $f_N({\bf x},1;{\bf x'},0)$ that define the behavior of ${\hat{\tilde{\beta}}}_z({\bf x},\xi_1;{\bf x'},\xi_2)$ are expressed in terms of the generating function of the probability of first arrival $f_N({\bf x})$, which is provided by Eq.~(\ref{candfgenerating}). In Appendix~\ref{twopintgen} we show that
\begin{equation}
    \begin{array}{lll}
         {\hat f}_z({\bf x},1;{\bf x'},0) & = 
         &  
         \frac{{\hat f}_z({\bf x})-{\hat f}_z({\bf x'}){\hat f}_z({\bf x-x'})}{1-{\hat f}_z({\bf x'-x}){\hat f}_z({\bf x-x'})}
         \\
         {\hat M}_z({\bf x},{\bf x'}) & =
         & 
         \frac{{\hat f}_z({\bf x'-x})-{\hat f}_z({\bf 0}){\hat f}_z({\bf x'-x})}{1-{\hat f}_z({\bf x'-x}){\hat f}_z({\bf x-x'})}
         \\
         {\hat T}_z({\bf x},{\bf x'}) & =
         & 
          \frac{{\hat f}_z({\bf 0})-{\hat f}_z({\bf x'-x}){\hat f}_z({\bf x-x'})}{1-{\hat f}_z({\bf x'-x}){\hat f}_z({\bf x-x'})}.
    \end{array}
    \label{twopointonepoint}
\end{equation}
Since the generating function of $\beta_N({\bf x},k_1;{\bf x'},k_2)$ is represented in terms of ${\hat f}_z({\bf x})$, ${\hat f}_z({\bf x'})$, ${\hat f}_z({\bf x-x'})$ and ${\hat f}_z({\bf x'-x})$, the summation over ${\bf x}$ and ${\bf x'}$ can be achieved in the $t\to\infty$ limit. Due to Eq.~(\ref{candfgenerating}) and the already mentioned fact that $\sum_{\bf x}{\hat c}_z({\bf x})=1/(1-z)$, the summation over all possible ${\bf x}$ and ${\bf x'}$ on the right hand side of Eq.~(\ref{twopointbeta02}) can be expanded in a power series over $1/(1-z)$. The Tauberian theorem~\cite{Weiss} states that the leading order in $t$ space is provided by the leading order of $1/(1-z)$ in the $z\to 1$ limit in $z$ space. It is clear that $\sum_{\bf x}\sum_{\bf x'}{\hat c}_z({\bf x}){\hat c}_z({\bf x'})=1/(1-z)^2$, but in Eq.~(\ref{twopointbeta02}) all the multiplications of generating functions of single point probabilities are of mixed origin, e.g. ${\hat c}_z({\bf x}){\hat c}_z({\bf x-x'})$, ${\hat c}_z({\bf x-x'}){\hat c}_z({\bf x'-x})$ and all other possibilities. Moreover substitution of Eq.~(\ref{twopointonepoint}) and Eq.~(\ref{candfgenerating})  in Eq.~(\ref{twopointbeta02}) shows that most of the multiplications will include more than two terms, e.g. ${\hat c}_z({\bf x}){\hat c}_z({\bf x-x'}){\hat c}_z({\bf x'-x})$. In Appendix~\ref{fouriecxz} we show that 
\begin{equation}
\sum_{\bf x}\sum_{\bf x'} {\hat c}_z({\bf x}){\hat c}_z({\bf x'-x})= \frac{1}{(1-z)^2}
\label{doublesumz01}
\end{equation}
for any case of transient RW (the roles of ${\bf x}$ and ${\bf x'}$ can be interchanged). Any other terms of the form $\sum_{\bf x}\sum_{\bf x'} {\hat c}_z({\bf x-x'}){\hat c}_z({\bf x'-x})$ or $\sum_{\bf x}\sum_{\bf x'} {\hat c}_z({\bf x}){\hat c}_z({\bf x}){\hat c}_z({\bf x'-x})$ (or generally multiplication of any number of terms greater than $2$) grow slower than $1/(1-z)^2$ when $z\to 1$ (see Appendix~\ref{fouriecxz}). This means that when expanding the denominator in Eq.~(\ref{twopointbeta02}) and utilizing Eq.~(\ref{twopointonepoint}), all the terms in the expansion, except the zero order, i.e. $1/[1-\xi_1{\hat T}_z({\bf x'},{\bf x})]
      [1-\xi_1{\hat T}_z({\bf x},{\bf x'})]$, will grow slower than $1/(1-z)^2$ after summation over ${\bf x}$ and ${\bf x'}$.
Then in the $z\to 1$ limit we use 
\begin{equation}
    \begin{array}{llll}
         {\hat f}_z({\bf x},1;{\bf x'},0) & = 
         &  
         {\hat f}_z({\bf x})
         &
         \\
         {\hat M}_z({\bf x},{\bf x'}) & =
         & 
         {\hat f}_z({\bf x'-x})-{\hat f}_z({\bf 0}){\hat f}_z({\bf x'-x})
         & \qquad (z\to 1)
         \\
         {\hat T}_z({\bf x},{\bf x'}) & =
         & 
          {\hat f}_z({\bf 0})
          &.
    \end{array}
    \label{zto1lim01}
\end{equation}
and the only relevant terms in the summation over ${\bf x}$ an ${\bf x'}$ are 
\begin{equation}
\begin{array}{ll}
\sum_{\bf x}\sum_{\bf x'}
{\hat{\tilde{\beta}}}_z({\bf x};\xi_1,{\bf x'};\xi_2)\underset{z\to 1}{\longrightarrow}
         \sum_{\bf x}\sum_{\bf x'}
         \frac{(1-{\hat f}_z({\bf 0}))^2}{1-z}
         \frac{\xi_1 \xi_2}{[1-\xi_1{\hat f}_z({\bf 0})]
      [1-\xi_2{\hat f}_z({\bf 0})]
      }
      &
         \left\{
                  {\hat f}_z({\bf x'})
     {\hat f}_z({\bf x-x'})
      \right.
         \\
        &
        \left.+
            {\hat f}_z({\bf x})
     {\hat f}_z({\bf x'-x})
           \right\}.
      \end{array}
      \label{sumbetazto1}
\end{equation}
Substituting the expression in Eq.~(\ref{doublesumz01}) and Eq.~(\ref{candfgenerating}) into Eq.~(\ref{sumbetazto1}) leads to 
\begin{equation}
\sum_{\bf x}\sum_{\bf x'}
{\hat{\tilde{\beta}}}_z({\bf x};\xi_1,{\bf x'};\xi_2)\underset{z\to 1}{\longrightarrow}
                  \frac{2(1-{\hat f}_z({\bf 0}))^4}{(1-z)^3}
         \frac{\xi_1 \xi_2}{[1-\xi_1{\hat f}_z({\bf 0})]
      [1-\xi_2{\hat f}_z({\bf 0})]
      },
    \label{sumbetazto1_01}
\end{equation}
and since $\sum_{k=1}^\infty \xi^k \left({\hat f}_z({\bf 0})\right)^{k-1}=\xi\big/[1-\xi{\hat f}_z({\bf 0})]$
\begin{equation}
\sum_{\bf x}\sum_{\bf x'}
{\hat{{\beta}}}_z({\bf x};k_1,{\bf x'};k_2)\underset{z\to 1}{\longrightarrow}
                  \frac{2(1-{\hat f}_z({\bf 0}))^4}{(1-z)^3}
         {\hat f}_z({\bf 0})^{k_1-1}
         {\hat f}_z({\bf 0})^{k_2-1}.
    \label{sumbetazto1_1}
\end{equation}
Eventually from Eq.~(\ref{sumbetazto1_1}) and Eq.~(\ref{secmomdef}) we obtain 
\begin{equation}
{\overline{\hat{S_\alpha^2}}}(z)\underset{z\to 1}{\longrightarrow} \frac{(1-{\hat f}_z({\bf 0}))^4}{{\hat f}_z({\bf 0})^2} 
\left\{Li_{-\alpha}({\hat f}_z({\bf 0})\right\}^2
\frac{2}{(1-z)^3}
    \label{secmomzform01}
\end{equation}
then according to the identity $\sum_{N=0}^\infty z^N N(N-1)=2z^2/(1-z)^3$, and the Tauberian theorem, the asymptotic behavior of ${\overline{S_\alpha^2}}(N)$ is
\begin{equation}
{\overline{{S_\alpha^2}}}(N)\sim \frac{(1-Q_0)^4}{Q_0^2} 
\left\{Li_{-\alpha}(Q_0)\right\}^2
N^2 \qquad N\to\infty.
    \label{secmomzform02}
\end{equation}
This relation shows that for any transient RW, in the large $N$ limit the second moment of $S_{\alpha}(N)$ converges to a square of the mean of $S_{\alpha}(N)$, i.e. 
\begin{equation}
    \frac{{\overline{S_{\alpha}(N)^2}}}
    {{\overline{S_{\alpha}(N)}}^2}
    \underset{N\to\infty}{\longrightarrow}1.
    \label{convergensmom}
\end{equation}

\begin{figure} 
    \centering
    \begin{subfigure}[b]{0.48\textwidth}
        \includegraphics[width=\textwidth]{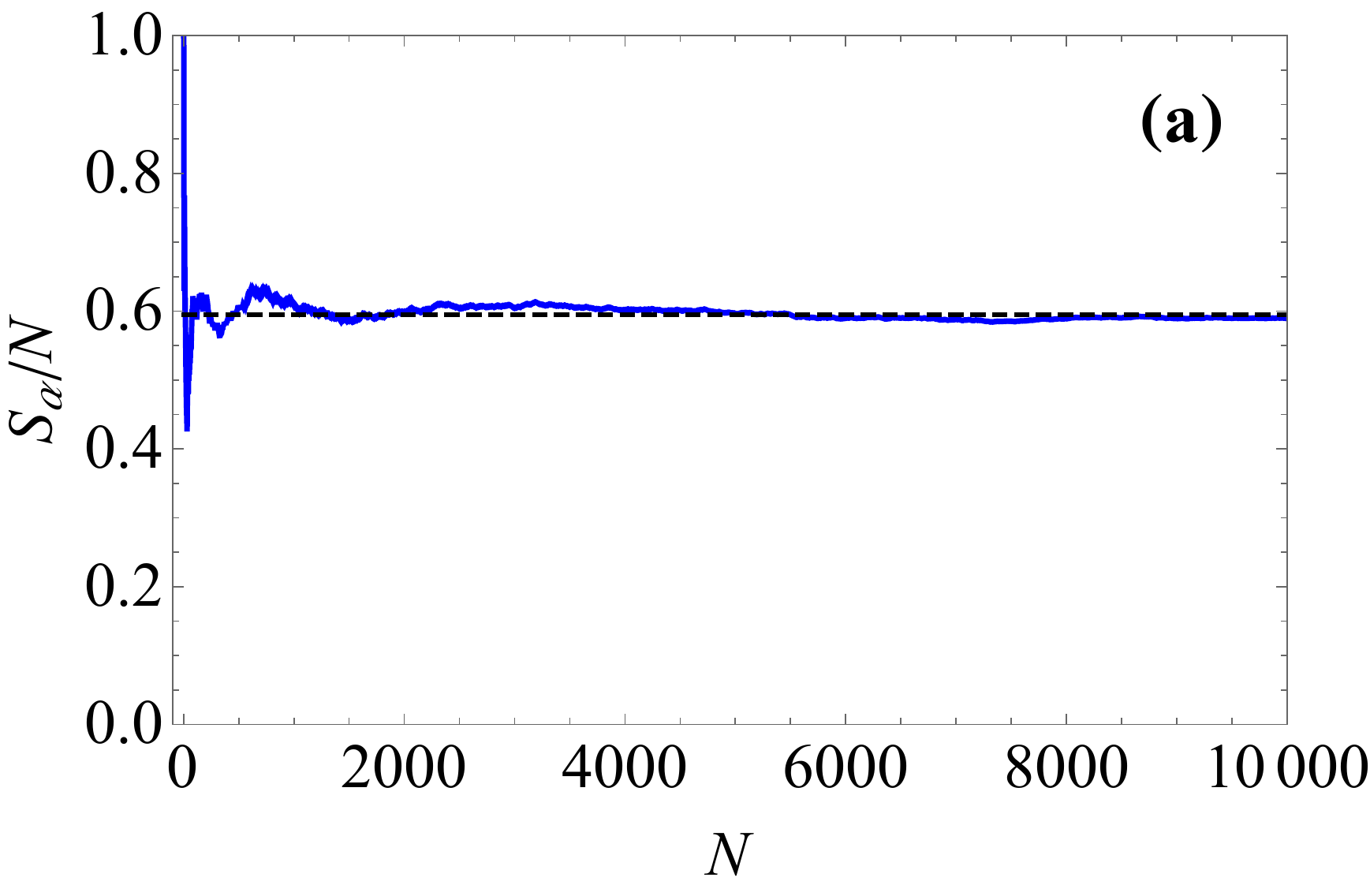}
           \end{subfigure}
 ~ 
    \begin{subfigure}[b]{0.48\textwidth}
        \includegraphics[width=\textwidth]{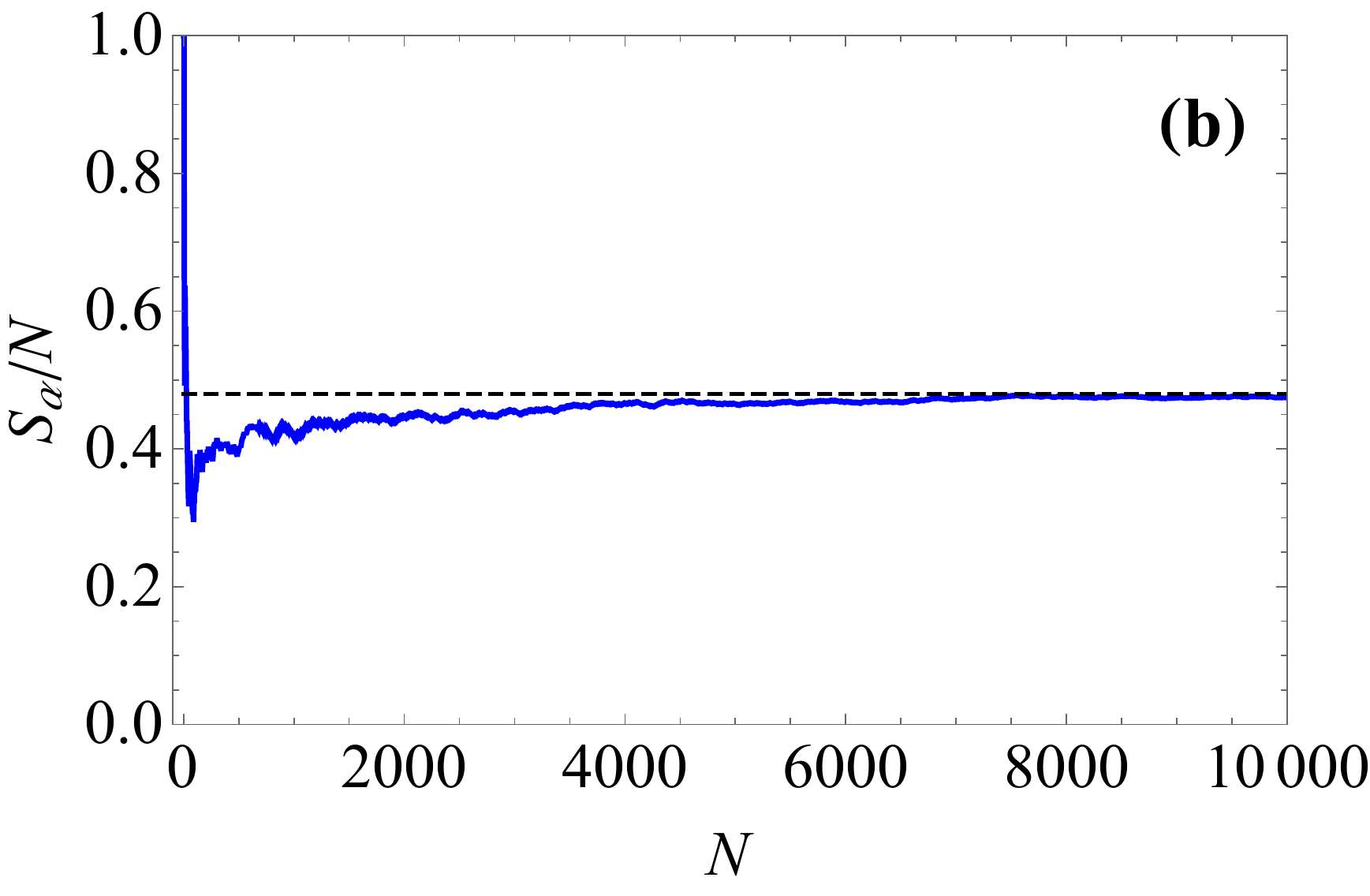}
    \end{subfigure}
    \caption{ Convergence of $S_\alpha(N)$ to $\Lambda N$. Both panels describe the behavior of $S_\alpha$ for a one dimensional RW with probability $0.7$ to make a step $+1$ and probability $0.3$ to make a step $-1$. The thick line in both panels are simulation results while the dashed line is the theoretical prediction of Eq.~(\ref{deltaconverge}) with $\Lambda$ provided by Eq.~(\ref{lambdaconst}). For both panels $Q_0=0.6$. Panel{\bf (a)} presents the case with $\alpha=0.5$ while panel {\bf (b)} the case with $\alpha=0.25$.  
    }
    \label{salphaconverge}
\end{figure}

\subsection{Convergence to a $\delta$-function}
\label{deltafunction}

We had shown that the distribution of $S_\alpha$ is such that in the $N\to\infty$ limit the square of the first moment converges to the second moment. The minimal value of $S_\alpha/N$ is $2(N/2)^\alpha /N$ that is achieved if the RW performed $N/2$ back and forward jumps between two sites. The maximal value of $S_\alpha /N$ is $1$, that is achieved if the RW never visited any site twice.
Since those two limits are achieved for a very specific trajectories of the RW the probability of the minimal and maximal values of $S_\alpha$ converges to $0$ in the $N\to\infty$ limit. 
For the random variable 
\begin{equation}
s=S_\alpha/N,
    \label{sdivNdef}
\end{equation}
the PDF $\lambda(s)$ is defined for $0\leq s \leq 1$ and $\lambda(s)\to 0$ when $s\to 0$, or $s\to 1$. 
Moreover the proven equivalence of ${\overline{S_\alpha^2}}$ and ${\overline{S_\alpha}}^2$ in the $N\to\infty$ limit means that
\begin{equation}
\left(\int_0^1s\lambda(s)\,ds\right)^2=\int_0^1\left(s\right)^2\lambda(s)\,ds\qquad N\to\infty.
    \label{jenseneq01}
\end{equation}
Since $\lambda(s)$ is a PDF and $\left(\dots\right)^2$ is a strictly convex function, Jensen inequality~\cite{Jensen} states that $\left(\int_0^1s\lambda(s)\,ds\right)^2\leq\int_0^1\left(s\right)^2\lambda(s)\,ds$ and the equality is achieved only when $s$ is constant, i.e. $\lambda(s)$ is a $\delta$-function. Then from Eq.~(\ref{salphaNlarge}) we obtain that
\begin{equation}
\lambda(s)\underset{N\to\infty}{\longrightarrow} \delta\left(s-\Lambda\right),
    \label{deltaconverge}
\end{equation}
where the constant $\Lambda$ is provided in Eq.~(\ref{lambdaconst}).
This result means that in the large $N$ limit the local time $S_\alpha$ and the number of jumps $N$ are equivalent up to a transformation $S_\alpha\to \Lambda N$. This result is presented in Fig.~\ref{salphaconverge}, where the random variable $S_\alpha(N)/N$ (obtained from numerical simulation) converges to a non-zero constant for large $N$. In the next section we utilize this result to establish the form of $P_{S_\alpha}({\bf x})$ and a simplified representation of the positional probability density function, i.e. $P({\bf x},t)$.

\section{Double subordination and the equivalence of CTRW and the transient QTM}
\label{doublesubordination}

\begin{figure} 
\centering
			\includegraphics[width=0.5\textwidth]{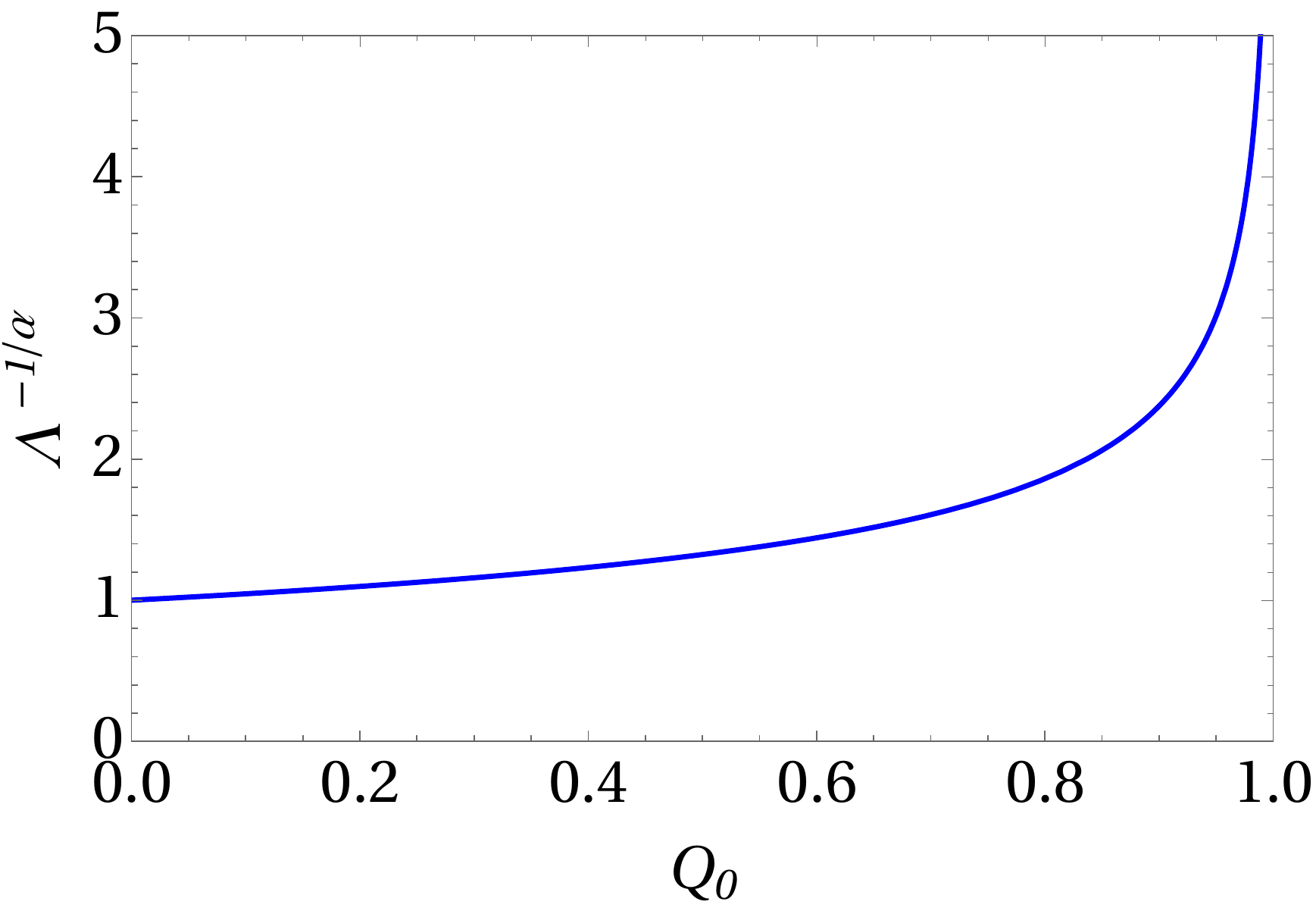}
\caption{
Behavior of the  $\Lambda=\frac{\left[1-Q_0\right]^2}{Q_0} Li_{-\alpha}(Q_0)$ (dependent pre-factor of the temporal transformation $t\to t/\Lambda^{1/\alpha}$) as a function of the return probability $Q_0$, for $\alpha=0.75$. The divergence for $Q_0\to 1$ signifies the limitation of this transformation strictly to the transient case.
}
    \label{qzeroplot}
\end{figure}

The PDF $P({\bf x},t)$, as it is presented by Eq.~(\ref{subordination02}), depends on  $P_{S_\alpha}({\bf x})$. 
The form of  $P_{S_\alpha}({\bf x})$ is obtained by using again the subordination approach where the local time $S_\alpha$ is subordinated to $N$ ,number of jumps performed, and the spatial process is given provided by $W_N{\bf x}$ - the PDF of regular RW, i.e.
\begin{equation}
    P_{S_\alpha}({\bf x})=\sum_{N=0}^\infty W_N({\bf x}) {\cal G}_{S_\alpha}(N,{\bf x}).
    \label{doublesub01}
\end{equation}
${\cal G}_{S_\alpha}(N,{\bf x})$ is the probability to perform $N$ steps before reaching ${\bf x}$ provided that the value of $S_\alpha$ is known. In the previous section we have shown that in the $N\to\infty$ limit the PDF of $s=S_\alpha/N$, i.e. $\lambda(s)$, is converging to $\delta(s-\Lambda)$. For $\lambda(s)$, $S_\alpha$ is the random variable and $N$ is the parameter. For ${\cal G}_{S_\alpha}$ $N$ is the random variable and $S_\alpha$ is the parameter. The convergence of $\lambda(s)$ to a $\delta$-function shows that in the $N\to\infty$ limit these two quantities are interchangeable and then for a transient RW 
\begin{equation}
    {\cal G}_{S_\alpha}(N,{\bf x})\underset{S_\alpha\to\infty}{\longrightarrow}
    \delta\left({S_\alpha-\Lambda N}\right),
    \label{gsalpharep}
\end{equation}
independent of the value of ${\bf x}$. 
The double subordination approach prescribes the disorder averaged PDF $\langle P({\bf x},t) \rangle$ the form 
\begin{equation}
\langle P({\bf x},t) \rangle=\sum_{S_\alpha}\sum_{N=0}^\infty 
W_N({\bf x}) {\cal G}_{S_\alpha}(N,{\bf x}) 
{\cal P}_t(S_\alpha)
    \label{doublesub02}
\end{equation}
where we used Eqs.(\ref{subordination01},\ref{doublesub01}). 
When taking the limit $t\to\infty$ the form of ${\cal P}_t(S_\alpha)$ in Eq.~(\ref{salphadist}) dictates that only large $S_\alpha$ need to be considered, and then according to Eq.~(\ref{gsalpharep}) only large $N$ are of interest, finally we obtain that
\begin{equation}
\langle P({\bf x},t) \rangle 
\sim\int_0^\infty W_{ N}({\bf x})
\frac{t\big/\Lambda^{1/\alpha}}{\alpha} N^{-1/\alpha-1}l_{\alpha,A,1}\left(\frac{t\big/\Lambda^{1/\alpha}}{N^{1/\alpha}}\right)\,dN
\qquad t\to\infty,
    \label{pxtformfin}
\end{equation}
where the transition to integration is the regular practice of the subordination technique~\cite{Bouchaud}. 
It is important to notice that in the case of continuous time random walk (CTRW)~\cite{Weiss} the particle experience each jump a new waiting time $\tau$, independent of the previous visitation even if itis currently located in a previously visited site. This makes the CTRW a kind of mean-filed approximation of the QTM and specifically, according to Eq.~(\ref{localtime}), for CTRW $S_\alpha=N$. Accordingly, only one level of subordination is needed and $P_{S_\alpha}({\bf x})$ is simply $W_N({\bf x})$ that leads to 
\begin{equation}
\langle P({\bf x},t) \rangle_{CTRW} 
\sim\int_0^\infty W_{ N}({\bf x})
\frac{t}{\alpha} N^{-1/\alpha-1}l_{\alpha,A,1}\left(\frac{t}{N^{1/\alpha}}\right)\,dN
\qquad t\to\infty.
    \label{pxtctrw}
\end{equation}
Comparison of Eq.~(\ref{pxtctrw}) and Eq.~(\ref{pxtformfin}) leads to 
\begin{equation}
\langle P({\bf x},t) \rangle_{QTM} \sim \langle P({\bf x},t/\Lambda^{1/\alpha})_{CTRW} \qquad t\to\infty,
    \label{equivalence}
\end{equation}
or simply said : the disorder averaged propagator of a transient QTM is equivalent to the propagator of CTRW taken at time $t/\Lambda^{1/\alpha}$. Eventually we proved that a simple transformation of time for CTRW 
\begin{equation}
t\to t\big/\Lambda^{1/\alpha}
    \label{timechange}
\end{equation}
makes this model sufficient to asymptotically represent the transient case of the QTM. Eq.~(\ref{equivalence}) states that for every situation that the propagator of CTRW can be computed~\cite{Barkai}, the propagator of QTM can be computed as well. The constant $\Lambda^{-1/\alpha}$ is provided by Eq.~(\ref{lambdaconst}) and displayed in Fig.~\ref{qzeroplot} for  $0\leq Q_0 <1$. 
This constant is positive and $>1$ for any $Q_0$. In the limit when $Q_0\to 1$, i.e. the approach to the recurrent case, $Li_{-\alpha}(Q_0)\sim (1-Q_0)^{-1-\alpha}$~\cite{Abramowitz} and $\Lambda^{-1/\alpha}\sim(1-Q_0)^{-(1-\alpha)/\alpha}$ diverges as $Q_0\to1$. This divergence  signifies the limitation of the presented result to the transient case $0\le Q_0 <1$. When $Q_0=0$ the QTM is exactly described by the CTW since the particle never returns to previously visited site, indeed in this case $\Lambda^{-1/\alpha}=1$. For any $0<Q_0 <1$ the constant is greater than $1$. This means that the QTM process is faster than CTRW, i.e. the two models attain the same PDFs but for QTM it is achieved on shorter time-scales. Such behavior can be attributed to the fact that CTRW never resamples previously visited traps (the disorder is annealed), while it is not true for QTM. Since CTRW never resamples previously visited traps it has a higher probability (when compared to QTM) to find deeper traps, which means that its propagation is going to be slower than QTM, on average.

\begin{figure} 
\centering
			\includegraphics[width=0.5\textwidth]{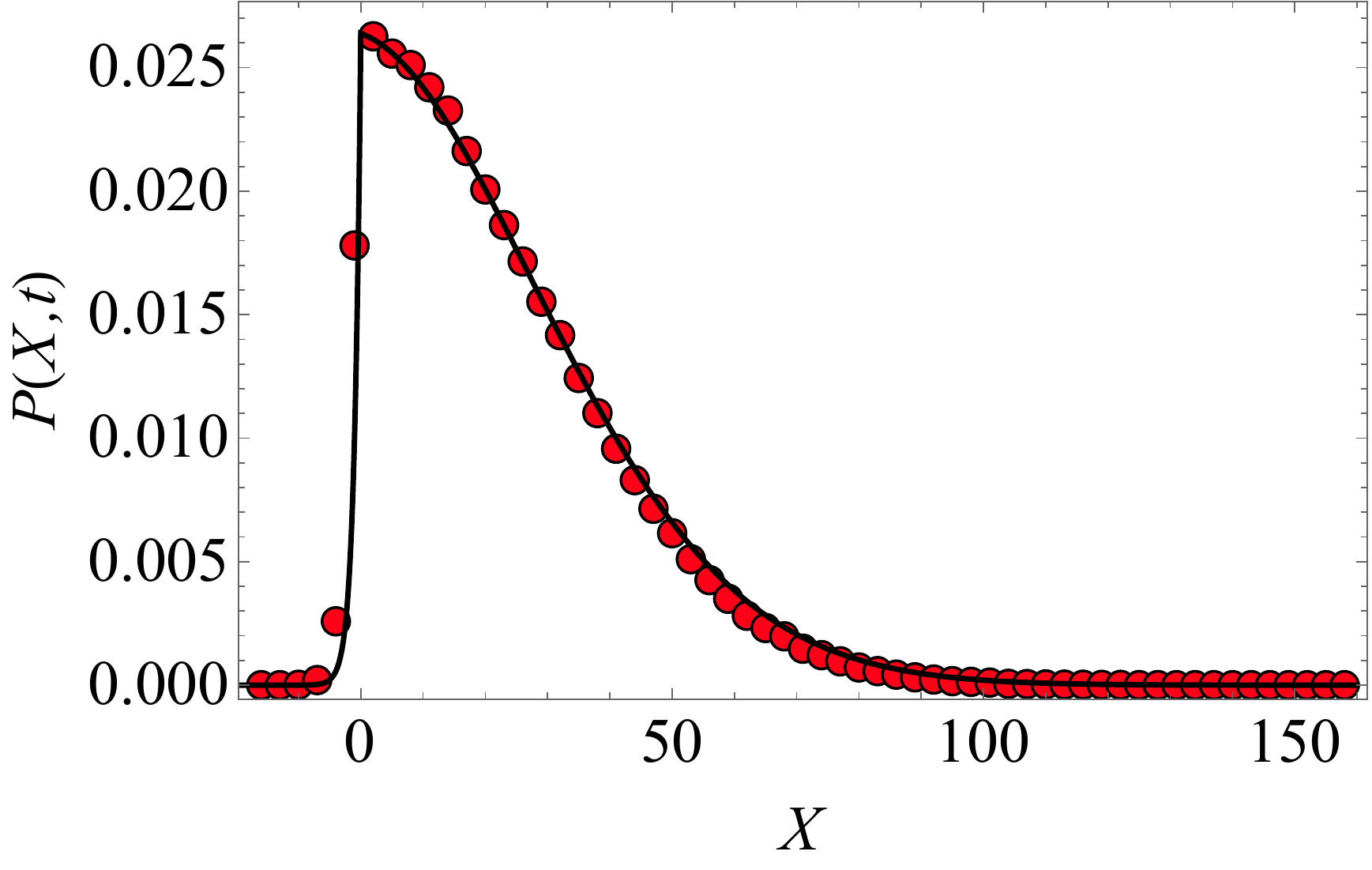}
\caption{
Comparison of the numerical simulation of the PDF for a 1d QTM with bias and theoretical predication of Eq.~(\ref{onedbiasedpxt}). The symbols is the numerical simulation while the thick line is theory without fitting parameters. The parameters of the case are : $A=1$, $q=0.7$, $\alpha=1/2$ and the spacing of the lattice is $1$.
}
    \label{pxtbiasfig}
\end{figure}

For the $1$-dimensional case of a biased RW on a simple lattice with constant spacing the $W_N( x)$ is a binomial distribution that is very well approximated by the Gaussian approximation
\begin{equation}
W_N(x)=\frac{1}{\sqrt{2\pi 4q(1-q)N}}e^{-\frac{\left(x-(2q-1)N\right)^2}{8q(1-q)N}}\qquad \left(N >> 1\right),
    \label{gauss1dbias}
\end{equation}
where $q$ is the probability to jump to the right one step on the lattice and $1-q$ is the probability to jump to the left. The return probability for this process is $Q_0=2(1-q)$, as proven in the next section. For several values of $\alpha$ the form of $l_{\alpha,A,1}$ is explicitly known\cite{Barkai}, specifically for $\alpha=1/2$, 
\begin{equation}
l_{1/2,1,1}(\eta)=\frac{1}{2\sqrt{\pi}}\eta^{-3/2}e^{-\frac{1}{4\eta}}.
    \label{lohalfdist}
\end{equation}
Then according to Eq.~(\ref{pxtformfin}), for the $1$-dimensional case the PDF is provided by
\begin{equation}
\begin{array}{ll}
\langle P(x,t) \rangle \sim  & \displaystyle
\int_0^\infty 
\frac{\sqrt{t} e^{-\frac{\left(x-(2q-1)N\right)^2}{8q(1-q)N}} }{\sqrt{2\pi^2 4q(1-q)N}}  \left(\frac{2(1-q)}{(2q-1)^2   Li_{-1/2}(2(1-q))}\right)^{-1}
\\
&
\exp\left[-\frac{N^2}{4t}\left(\frac{2(1-q)}{(2q-1)^2   Li_{-1/2}(2(1-q))}\right)^{2}\right]
\,dN\qquad (t\to\infty).
\end{array}
    \label{onedbiasedpxt}
\end{equation}
In Fig.~\ref{pxtbiasfig} we perform a comparison between a numerical simulation of the QTM and the theoretical result of Eq.~(\ref{onedbiasedpxt}). The comparison is performed for $t=10^3$ and it is excellent for this finite time.

\subsection{Moments of the QTM and non-linear response}
\label{nonlinresp}

The explicit form of the disorder averaged PDF, expressed by Eq.~(\ref{pxtformfin}), permits evaluation of different moments $\langle {\bf x}^\mu \rangle$. Indeed,  the approximation works for a regime when the the measurement time is sufficiently large and many jumps have been performed. In this limit the probability density $W_N({\bf x})$ attains the Gaussian form and all the moments $\int |{\bf x}|^\mu W_N({\bf x})d\,{\bf x}$ can be easily computed~\cite{Winkelbauer}. Generally we can say that  
\begin{equation}
\int |{\bf x}|^\mu W_N({\bf x})d\,{\bf x} = B_\mu N^{\gamma_\mu}. 
\label{gammamudef}
\end{equation}
The constant $B_\mu$ depends on the power $\mu$ and the lattice  that determine the properties of the Gaussian approximation, i.e. second moment and the mean of the Gaussian distribution. Then according to Eq.~(\ref{pxtformfin}) the $\mu$th moment $\langle |{\bf x}|^\mu \rangle$ is  provided by $\int_0^\infty (B_\mu t\big/\Lambda^{1/\alpha}\alpha)N^{\gamma_\mu-1-1/\alpha}l_{\alpha,A,1}\left(t/(\Lambda N)^{1/\alpha}\right)dN$. Since, $\int_0^\infty y^q l_{\alpha,A,1}(y)dy=A^{q/\alpha}\Gamma[1-q/\alpha]/\Gamma[1-q]$ (for $q/\alpha<1$)~\cite{Barkai} the expression for the moments of ${\bf x}$ takes the form
\begin{equation}
\langle |{\bf x}|^\mu \rangle= \frac{\Gamma[1+\gamma_\mu]}{A^{\gamma_\mu}\Gamma[1+\alpha\gamma_\mu]}\frac{B_\mu}{\Lambda^{\gamma_\mu}}
t^{\alpha\gamma_\mu}.
    \label{momentexpression}
\end{equation}
The constants $\gamma_\mu$, $B_\mu$ and $Q_0$ depend only on the lattice dimension  and the type of the RW on top of this lattice. 

Of a special interest is the behavior of the first moment when an external force is applied, i.e. response of the system to a bias. In the QTM model the force is applied in such a way that it is not affecting the dwell times $\tau_{\bf x}$ but rather determines the transition probabilities between different locations~\cite{Bertin02,MonthusSec,Deborah01}. When the imposed external force $F_0$ is sufficiently weak the transition probabilities $p({\bf x - x'})$ should be proportional to $\exp(F_0({\bf x-x'})/2k_BT)$ for transition from ${\bf x'}$ to ${\bf x}$, and to $\exp(-F_0({\bf x-x'})/2k_BT)$ for the reverse transition. Here we assume that the force is constant and applied in the direction of ${\bf x-x'}$, otherwise one needs to use the projection of the force in the ${\bf x-x'}$ direction. Since we are interested only in the limit of weak force it is possible to expand the exponential up to first order in $F_0$. In the case of a simple binomial RW on top of a $1$-dimensional lattice the probability $q$ to perform a jump to the  right will be $q=\frac{1}{2}(1+F_0a/2k_BT)$ and the probability to jump to the left $1-q=\frac{1}{2}(1-F_0a/2k_BT)$, where $a$ is the lattice spacing. For dimensions $d\geq 2$ similar expansion will take place, the only difference is that $F_0$ will be multiplied by some $\cos(\theta)$ where $\theta$ is the appropriate angle between the direction of the force and local axis of the lattice. 
The presence of the force affects not only the constant $B_\mu$ in Eq.~(\ref{momentexpression}) but also the constant $\Lambda$ by the means of $Q_0$. 
Of special interest is the one-dimensional case. For $d=1$ $Q_0$, without the presence of external force, is $1$~\cite{Weiss}. 
When external small external force  $F_0$ is added $Q_0$ is decreased but still attains values in the vicinity of $1$ and consequently (due to the form of $\Lambda$ in Eq.~(\ref{lambdaconst})) contributes to a non-trivial dependence on the force of the first moment.

The first moment of the one dimensional case with a presence of a weak force $F_0$ is the case of traps on a one a simple one-dimensional lattice with probabilities $q=\frac{1}{2}(1+F_0a/2k_BT)$ to jump to the right and $1-q$ to jump to the left. 
For the spatial process $W_N(x)$ this is the case of a binomial random walk and thus for sufficiently large $N$ the Gaussian limit is attained
\begin{equation}
W_N(x)\sim\frac{\exp\left[-\frac{\left(x-(2q-1)N\right)^2}{8q(1-q)N}\right]}
{\sqrt{8\pi q(1-q)N}}
    \label{binomial1dspat}
\end{equation}
and 
\begin{equation}
\int_{-\infty}^{\infty} x W_N(x)\,dx = (2q-1) N
    \label{binom1dmoment}
\end{equation}
meaning that $B_1=2q-1$ and $\gamma_1=1$. Eq.~(\ref{binom1dmoment}) describes the linear response to the external force for the spatial part of the QTM.
The return probability $Q_0\underset{z\to 1}{\to}\sum_{N=0}^\infty f_N({\bf 0})={\hat f}_{z}({\bf 0})$ is provided by Eq.~(\ref{candfgenerating}) while the 
Fourier transform of the jump probability $p({\bf x})$ is ${\overline p({\bf k})}=\sum_{\bf x} e^{i({\bf k \dot x})}p({\bf x})$ dictates the form of ${\hat c}_z({\bf 0})$ for dimension $d$~\cite{Weiss}
\begin{equation}
  {\hat c}_z({\bf 0})=\frac{1}{(2\pi)^d}\int_{-\pi}^{\pi}\dots\int_{-\pi}^{\pi}\frac{1}{1-z{\overline p({\bf k})}}\,d^d{\bf k}.   
\label{cxgenerating}
\end{equation}
For $d=1$, ${\overline{p}}(k)=q\exp(ik)+(1-q)\exp(-ik)$ and Eq.~(\ref{cxgenerating}) is 
\begin{equation}
{\hat c}_z({\bf 0})=\frac{1}{2\pi}\int_{-\pi}^{\pi}\frac{1}{1-z(q\exp(i k)+(1-q)\exp(-i k)}\,dk,
\label{cxgendeq1}
\end{equation}
by changing the variable to $y=\exp(ik)$ the integral in Eq.~(\ref{cxgendeq1}) is transformed into
\begin{equation}
{\hat c}_z({\bf 0})=\frac{1}{2\pi i}\oint_{|y|=1}\frac{1}{y-zq y^2-z(1-q)}\,dy.
    \label{cxgendeq1_01}
\end{equation}
For any $z<1$ the two solutions of $-zq y^2+y-z(1-q)=0$ are located on the real line while one of them is for $y>1$ and the other is is for $y<1$. 
This means that the integral depends on the presence of one single pole $\forall z<1$. This pole is located at $y=(1-q)/q$ for $z=1$ and the integral in Eq.~(\ref{cxgendeq1_01}) in the $z\to 1$ limit is 
\begin{equation}
{\hat c}_0({\bf 0})=\frac{1}{2q-1}.
    \label{cxgebdeq1_02}
\end{equation}
Then according to Eq.~(\ref{candfgenerating}) for $d=1$ the probability to return to the starting point, given the process is biased (i.e. $q>1/2$), is 
\begin{equation}
Q_0=2(1-q).
    \label{qzerodeq1}
\end{equation}
Finally, according to Eqs.~(\ref{momentexpression},\ref{binom1dmoment},\ref{qzerodeq1}) and Eq.~(\ref{lambdaconst}) we obtain that
\begin{equation}
\langle x(t) \rangle \underset{t\to\infty}{\sim}\frac{1}{A\Gamma[1+\alpha]}
\frac{2(1-q)}{(2q-1) Li_{-\alpha}\left[2(1-q)\right]} t^\alpha,
    \label{xmoment1d01}
\end{equation}
and when explicitly writing the the probability $q=\frac{1}{2}(1+F_0a/2k_BT)$ and the fact that the spacing of the lattice is $a$, $\langle x(t) \rangle$ is transformed into 
\begin{equation}
\langle x(t) \rangle \underset{t\to\infty}{\sim}\frac{a}{A\Gamma[1+\alpha]}
\frac{[1-F_0a/2k_BT]}{(F_0a/2k_BT) Li_{-\alpha}\left[1-F_0a/2k_BT\right]} t^\alpha.
    \label{xmoment1d01_xa01}
\end{equation}
For small $F_0\to 0$ we use the asymptotic relation $Li_{-\alpha}(1-y)\sim \Gamma[1+\alpha]y^{-\alpha-1}$~\cite{Abramowitz} and obtain the non-linear response to externally applied small force 
\begin{equation}
\langle x(t) \rangle \underset{t\to\infty}{\sim}\frac{a}{A\Gamma[1+\alpha]^2}
\left(\frac{F_0a}{2k_BT} \right)^\alpha t^\alpha.
    \label{xmoment1d01_xa02}
\end{equation}

\begin{figure} 
\centering
			\includegraphics[width=0.6\textwidth]{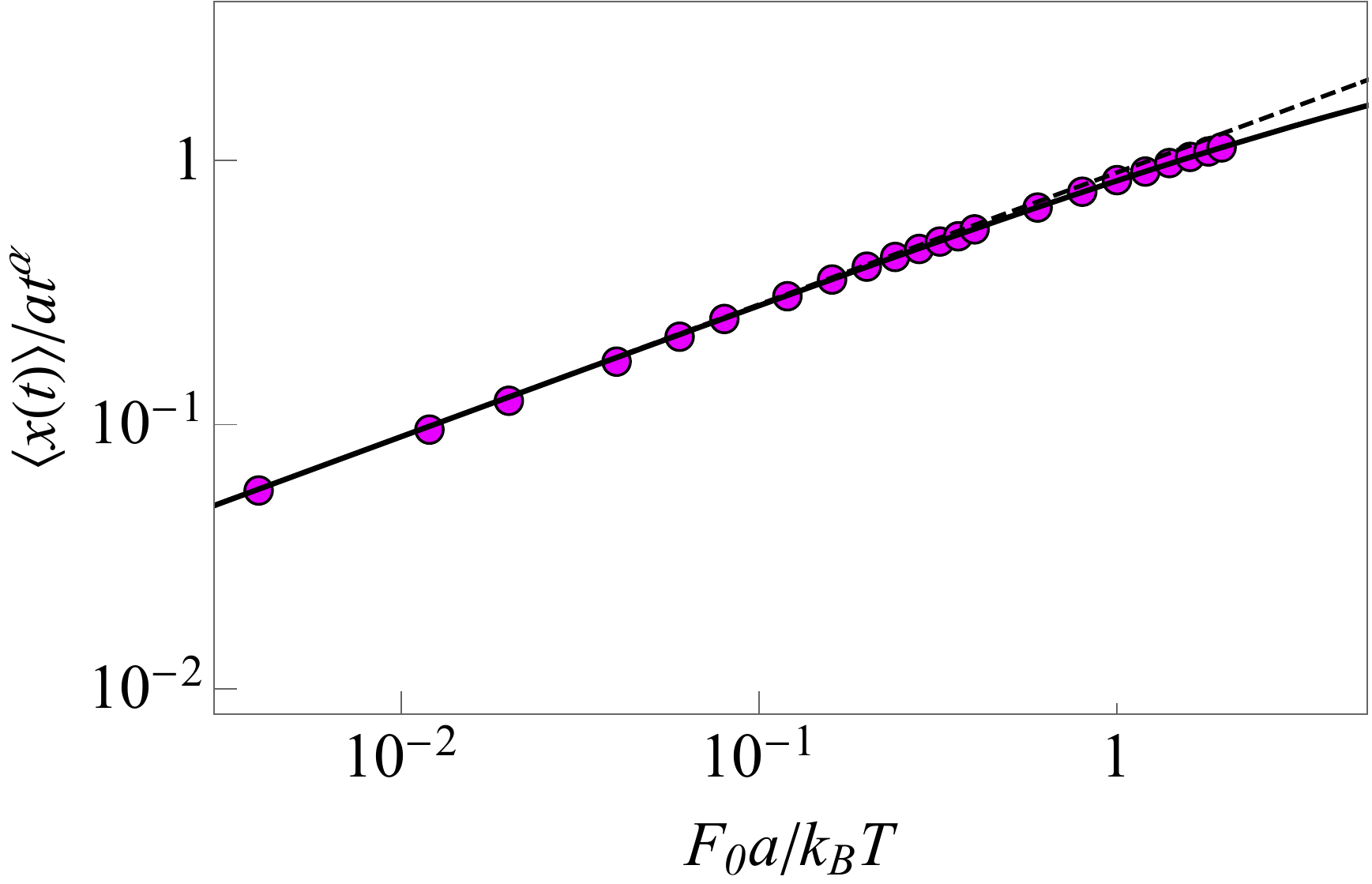}
\caption{
Comparison of the numerical simulation of the first moment $\langle x(t)\rangle$ for a 1d QTM with a presence of external force $F_0$ and theoretical predication of Eqs.~(\ref{xmoment1d01_xa01},\ref{xmoment1d01_xa02}). The symbols describe the results of the numerical simulation with $t=10^8$, $\alpha=1/2$, $a=1$ and $A=1$. The thick line is theory as described by Eq.~(\ref{xmoment1d01_xa01}) without fitting parameters and the dashed line is the prediction of Eq.~(\ref{xmoment1d01_xa02}). For sufficiently small $F_0a/k_BT$ the two theoretical results coincide.
}
    \label{biasedforcefig}
\end{figure}
A convincing comparison between the analytical results of Eqs.~(\ref{xmoment1d01_xa01},\ref{xmoment1d01_xa02}) and numerical simulation is presented in Fig.~\ref{biasedforcefig}. 
It is clear from the figure that both theoretical result due to Eq.~(\ref{xmoment1d01_xa01}) and Eq.~(\ref{xmoment1d01_xa02}) coincide for sufficiently small external force $F_0$.

The behavior of the first moment for small forces, as described by Eq,~(\ref{xmoment1d01_xa02}) does not satisfy linear response. 
The response to external force is anomalous and the force enters the equation with an exponent $\alpha<1$. 
This behavior for a $1$-dimensional biased QTM was previously predicted by using scaling analysis~\cite{Bouchaud,Bertin02} and also obtained by exploitation of the Renormalization Group techniques in the limit of low temperatures~, i.e $\alpha\to 0$~\cite{MonthusSec}. The non-linear response is present only due to the strong disorder and the quenched nature of the disorder. For the annealed case with power-law waiting times the response is linear~\cite{Bouchaud}. From the treatment of the $1$-dimensional case it becomes clear that the non-linearity appears solely due to presence of $\Lambda$ in the denominator of Eq.~(\ref{momentexpression}). According to Eq.~(\ref{lambdaconst}) $\Lambda$ depends on $Q_0$ in a non-trivial fashion. When a small external force is present it alters the probability of return $Q_0$. Of special interest are the cases where $Q_0=1$ when $F_0=0$. Addition of small $|F_0|$ will decrease $Q_0$ and introduce a non-linear contribution due to the divergence $\Lambda$ in the limit of $Q_0\to1$. For the cases where $Q_0<1$ even when the external force is non-present, addition of a non-zero external force slightly decreases $Q_0$ that is translated to a small change in $\Lambda$ and the linear response is not affected. It is then according to classical result of P{\'o}lya~\cite{Weiss}, the non-linear response is to be expected for $d=1,2$ while for any higher dimension the strong quenched disorder will not alter the linear response to external field.

\section{Summary}
\label{summary}

The properties of transport in the QTM have been extensively explored over the years. In this manuscript we provided an explicit mapping between the transient cases of QTM and the widely used CTRW. This result allows  to generalize any result that is known for the CTRW to the case of QTM. Immediate applications include, first-passage properties~\cite{Redner}, super-diffusive fluctuations for anomalous transport~\cite{Lindenberg,Voituriez}, representation by the means of fractional equations~\cite{Klafter}, large deviation properties~\cite{BarkaiBurov20} and many more.  The non trivial dependence of the mapping on the probability to return to the origin, $Q_0$, implies that we should expect very important differences between the QTM and CTRW for low dimensions even when the process is transient. Like the  existence of non-linear response to externally applied field that was calculated for the QTM and is absent for CTRW. 
The developed theoretical framework of double subordination and two-point probabilities have merit on their own. We hope that these methods will help in addressing the recurrent case of QTM. Finally we would like to notice that existence of explicit mappings between the QTM and other models of transport in disordered media, such as the barrier model~\cite{Sollich}, can allow to address the general case of transport in a random-potential landscape~\cite{SokolovCamb}.

{\bf Acknowledgments:} This work was supported by the Pazy foundation grant 61139927. I thank D.A. Kessler for fruitful discussions.

\section{Appendix}

\subsection{Additional terms of ${\hat{\psi}}(u)$}
\label{sbetaproof}

In Section~\ref{loctime} it was shown that when the expansion of ${\hat \psi}(u)$ is of the form ${\hat \psi}(u)\sim 1-Au^\alpha$, Eq.~(\ref{etalaplacefnl}) holds. 
Here we show that additional terms in the expansion, i.e. ${\hat \psi}(u)\sim 1-Au^\alpha+Bu^\beta$ with $\beta>\alpha$, won't change this equation when $S_\alpha \to \infty$.
In such a case
\begin{equation}
\langle e^{-u\eta}\rangle = \displaystyle \prod_{\bf x} \left( 1-\frac{n_{\bf x}^\alpha}{S_\alpha}Au^\alpha +\frac{n_{\bf x}^\beta}{S_\alpha^{\beta/\alpha}}Bu^\beta\right)
\label{betprooffull01}    
\end{equation}
and the multiplication will produce the terms mentioned in Sec.~\ref{loctime} and also terms of the form 
$\sum_{\bf x}n_{\bf x}^\beta B u^\beta/S_\alpha ^{\beta/\alpha}$, 
$\sum_{\bf x}\sum_{\bf x'}n_{\bf x}^\alpha n_{\bf x'}^\beta A B u^{\alpha+\beta}/S_\alpha^{1+\beta/\alpha}$, $\sum_{\bf x}\sum_{\bf x'}n_{\bf x}^\beta n_{\bf x'}^\beta B^2 u^{2\beta}/S_\alpha^{2\beta/\alpha}$ etc. 
Since $\sum_{\bf x} n_{\bf x}^\beta=S_\beta$, the behavior of the the term $\sum_{\bf x}n_{\bf x}^\beta B u^\beta/S_\alpha ^{\beta/\alpha}$ is dictated by the ratio $S_\beta/S_\alpha ^\beta/\alpha$. For the transient case , i.e presence of bias or $d>2$, we have shown in Sec.~\ref{salphaSec} that ${\overline S_\alpha}\sim \Lambda N$ when $N\to\infty$. 
This means that in the limit of many jumps, $N\to\infty$, the ratio $S_\beta/S_\alpha ^\beta/\alpha$ is decaying like $N^{-\frac{\beta}{\alpha}+1}$, ($\beta>\alpha$). 
Therefore, all the terms that are not of the form $\left(\sum_{\bf x} \frac{n_{\bf x}^\alpha}{S_\alpha} A u^\alpha\right)^j$ will decay to $0$ in the $N\to\infty$ limit. 
We can then state that only the two first terms in the expansion of ${\hat \psi}(u)$ ($1-Au^\alpha$) are needed.   

\subsection{Generating functions of two-point probabilities}
\label{twopintgen}

In Sec.~\ref{secondsalpha} three two-point probabilities were crucial for the behavior of $\beta_N({\bf x},k_1;{\bf x'},k_2)$ : {\bf I} $f_N({\bf x},1;{\bf x'},0)$, {\bf II} $M_N({\bf x},{\bf x'})$ and {\bf III} $T_N({\bf x},{\bf x'})$. 

The probability $f_N({\bf x},1;{\bf x'},0)$ is the probability to start at point ${\bf 0}$ and after $N$ steps to reach the point ${\bf x}$ for the first time, without visiting ${\bf x'}$ even once. So from all the possibilities to reach ${\bf x}$ for the first time after $N$ we must subtract those where the point ${\bf x'}$ was visited at-least once (before reaching ${\bf x}$), i.e.
\begin{equation}
f_N({\bf x},1;{\bf x'},0)=f_N({\bf x}) - \sum_{l=0}^N f_l({\bf x'},1;{\bf x},0)f_{N-l}({\bf x-x'}),
    \label{app_fngen01}
\end{equation}
where $f_N({\bf x})$ is the first-passage probability defined in Eq.~(\ref{candfgenerating}). The translational invariance of the lattice was utilized. According to Eq.~(\ref{app_fngen01}) the $z$-transform of $f_N({\bf x},1;{\bf x'},0)$ is 
\begin{equation}
   {\hat f}_z({\bf x},1;{\bf x'},0)={\hat f}_z({\bf x}) - {\hat f}_z({\bf x'},1;{\bf x},0){\hat f}_z({\bf x-x'}). 
    \label{app_fngen02}
\end{equation}
By switching the places of ${\bf x}$ and ${\bf x'}$ in Eq.~(\ref{app_fngen01}) and performing a $z$-transform we obtain
\begin{equation}
   {\hat f}_z({\bf x'},1;{\bf x},0)={\hat f}_z({\bf x'}) - {\hat f}_z({\bf x},1;{\bf x'},0){\hat f}_z({\bf x'-x}). 
    \label{app_fngen03}
\end{equation}
Substitution of Eq.~(\ref{app_fngen03}) into Eq.~(\ref{app_fngen02}) leads to an expression for ${\hat f}_z({\bf x},1;{\bf x'},0)$ in terms of a generating function of $f_N({\bf x})$
\begin{equation}
{\hat f}_z({\bf x},1;{\bf x'},0)=
\frac{{\hat f}_z({\bf x})-{\hat f}_z({\bf x'}){\hat f}_z({\bf x-x'})}{1-{\hat f}_z({\bf x'-x}){\hat f}_z({\bf x-x'})}.
    \label{app_fngen04}
\end{equation}

The probability $M_N({\bf x},{\bf x'})$ is the probability to start at ${\bf x}$ and after $N$ steps to reach ${\bf x'}$ for the first time, without returning to ${\bf x}$ on the way. Due to translational invariance of the lattice $M_N({\bf x},{\bf x'})$ is expressible in terms of $f_N({\bf x},1;{\bf x'},0)$, i.e. $M_N({\bf x},{\bf x'})=f_N({\bf x'-x},1;{\bf 0},0)$. Then according to Eq.~(\ref{app_fngen04}) the generating function of $M_N({\bf x},{\bf x'})$ is 
\begin{equation}
{\hat M}_z({\bf x},{\bf x'})=
\frac{{\hat f}_z({\bf x'-x})-{\hat f}_z({\bf 0}){\hat f}_z({\bf x'-x})}{1-{\hat f}_z({\bf x'-x}){\hat f}_z({\bf x-x'})}.
    \label{app_mgen01}
\end{equation}

The probability $T_N({\bf x},{\bf x'})$ is the probability to return to ${\bf x}$ after $N$ steps without visiting ${\bf x'}$ on the way. Once again the translational invariance of the lattice allows to utilize $f_N({\bf x},1;{\bf x'},0)$ and hence $T_N({\bf x},{\bf x'})=f_{N}({\bf 0},1;{\bf x-x'},0)$. Then according to Eq.~(\ref{app_fngen04}), the generating function of $T_N({\bf x},{\bf x'})$ is provided by
\begin{equation}
{\hat T}_z({\bf x},{\bf x'})  =
                   \frac{{\hat f}_z({\bf 0})-{\hat f}_z({\bf x'-x}){\hat f}_z({\bf x-x'})}{1-{\hat f}_z({\bf x'-x}){\hat f}_z({\bf x-x'})}.
    \label{app_tgen01}
\end{equation}

\subsection{Properties of $c_N({\bf x})$ and summation over all lattice points}
\label{fouriecxz}

The probability to find the particle at position ${\bf x}$ after $N$ steps (when starting at ${\bf 0}$), $c_N({\bf x})$ is normalized, i.e. $\sum_{\bf x}c_N({\bf x})=1$, where the summation is over all possible lattice points. This leads to the following relation
\begin{equation}
\sum_{\bf x} c_N({\bf x}) e^{i{\bf a\cdot x}}
\underset{{\bf a}\to{\bf 0}}{\longrightarrow}1
    \label{afr_single01}
\end{equation}
and consequently for the generating function ${\hat c}_z({\bf x})=\sum_{N=0}^\infty z^N c_N({\bf x})$
\begin{equation}
\sum_{\bf x} {\hat c}_z({\bf x}) e^{i{\bf a\cdot x}}
\underset{{\bf a}\to{\bf 0}}{\longrightarrow}\frac{1}{1-z}.
    \label{afr_single02}
\end{equation}
For the single jump probability $p({\bf x})$ the characteristic function is defined as 
${\hat p}({\bf a})=\sum_{x_1}\sum_{x_2}\dots\sum_{x_d}p({\bf x})e^{i{\bf a\cdot x}}$, where ${\bf x}=(x_1,x_2,\dots,x_d)$ are all possible single steps on the lattice. Since all the jumps of the RW on the lattice are independent, $\sum_{\bf x} c_N({\bf x})e^{i{\bf a\cdot x}}=\left({\hat p}({\bf a})\right)^N$ and according to Eq.~(\ref{afr_single02})
\begin{equation}
\sum_{\bf x} {\hat c}_z({\bf x}) e^{i{\bf a\cdot x}}=
\frac{1}{1-z{\hat p}({\bf a})}
\underset{{\bf a}\to{\bf 0}}{\longrightarrow}\frac{1}{1-z}.
    \label{afr_single03}
\end{equation}

According to Eq.~(\ref{afr_single03}) the double sum $\sum_{\bf x}\sum_{\bf x'} {\hat c}_z({\bf x}) {\hat c}_z({\bf x'})$ is simply
\begin{equation}
\sum_{\bf x}\sum_{\bf x'} {\hat c}_z({\bf x}) {\hat c}_z({\bf x'})=\underset{{\bf a}\to{\bf 0}}{\lim}\sum_{\bf x}\sum_{\bf x'} {\hat c}_z({\bf x}) {\hat c}_z({\bf x'})e^{i{\bf a\cdot x}}e^{i{\bf a\cdot x'}}=
\underset{{\bf a}\to{\bf 0}}{\lim}\frac{1}{\left(1-z{\hat p}({\bf a})\right)^2}=\frac{1}{(1-z)^2}.
    \label{afr_double01}
\end{equation}
This result is simply extended to the case of 
$\sum_{\bf x}\sum_{\bf x'} {\hat c}_z({\bf x}) {\hat c}_z({\bf x'-x})$. Indeed, 
\begin{equation}
\sum_{\bf x}\sum_{\bf x'} {\hat c}_z({\bf x}) {\hat c}_z({\bf x'-x})e^{i{\bf a\cdot x}}e^{i{\bf a\cdot x'}}=\sum_{\bf x}{\hat c}_z({\bf x})e^{i2{\bf a\cdot x}}
\sum_{\bf x'}{\hat c}_z({\bf x'-x})e^{i{\bf a\cdot(x'-x)}},
    \label{afr_double02}
\end{equation}
due to translational invariance the right hand side of Eq.~(\ref{afr_double02}) equals to 
$\frac{1}{1-z{\hat p}(2{\bf a})}\frac{1}{1-z{\hat p}({\bf a})}$ and we obtain
\begin{equation}
\sum_{\bf x}\sum_{\bf x'} {\hat c}_z({\bf x}) {\hat c}_z({\bf x'-x})e^{i{\bf a\cdot x}}e^{i{\bf a\cdot x'}}=\underset{{\bf a}\to{\bf 0}}{\lim}\frac{1}{1-z{\hat p}(2{\bf a})}\frac{1}{1-z{\hat p}({\bf a})}=\frac{1}{(1-z)^2}.
    \label{afr_double03}
\end{equation}
Sums of terms of the form ${\hat c}_z({\bf x'}){\hat c}_x({\bf x-x'})$ produce similar result. Generally speaking, when the arguments of ${\hat c}_z(\dots){\hat c}_z(\dots)$ cover all possible  points $({\bf x},{\bf x'})$ of the $2d$ lattice, the double summation will provide the result $1/(1-z)^2$.

We turn now to calculation of sums of the form 
$\sum_{\bf x}\sum_{\bf x'} {\hat c}_z({\bf x})
{\hat c}_z({\bf x'-x}){\hat c}_z({\bf x-x'})$. For this case the behavior of $\sum_{\bf x'}{\hat c}_z({\bf x'-x}){\hat c}_z({\bf x -x'})e^{i{\bf a\cdot x'}}$ must be inspected. 
According to the convolution theorem
\begin{equation}
\sum_{\bf x'}{\hat c}_z({\bf x'}){\hat c}_z(-{\bf x'})e^{i{\bf a\cdot x'}}=\left(\frac{1}{2\pi}\right)^d
\int_{-\pi}^{\pi}\dots\int_{-\pi}^{\pi}
\frac{1}{1-z{\hat p}({\bf b})}\frac{1}{1-z{\hat p}({\bf b-a})}d^d{\bf b},
    \label{afr_triple01}
\end{equation}
where $d^d{\bf b}$ is $db_1\,db_2\dots db_d$. When the ${\bf a}\to{\bf  0}$ limit is taken, the integrand on the right hand side of Eq.~(\ref{afr_triple01}) is simply $1\big/(1-z{\hat p}({\bf b}))^2$. 
Moreover, the asymptotic limit of $N\to\infty$ is translated as the $z\to 1$ limit in the $z$ space. In this limit the main contribution to the integral in Eq.~(\ref{afr_triple01}) is from the values of ${\bf b}$ that are in the vicinity of ${\bf 0}$, since ${\hat p}({\bf 0})=1$ and the integrand converges to $1/(1-z)^2$. We concentrate on two types of ${\hat p}({\bf b})$ expansions in the vicinity of ${\bf b = 0}$. The first type is a linear case
\begin{equation}
{\hat p}({\bf b})\sim 1+i{\bf b \cdot B}\qquad {\bf b}\to 0.
    \label{afr_tripexp01}
\end{equation}
This is the case of a RW with a bias in the ${\bf B}$ direction. Then 
\begin{equation}
\left(\frac{1}{2\pi}\right)^d
\int_{-\pi}^{\pi}\dots\int_{-\pi}^{\pi}
\frac{1}{\left(1-z{\hat p}({\bf b})\right)^2}d^d{\bf b}
\underset{z\to 1}{\sim}
\left(\frac{1}{2\pi}\right)^d
\int_{-\pi}^{\pi}\dots\int_{-\pi}^{\pi}
\frac{1}{\left(1-z(1+i{\bf b\cdot B})\right)^2}d^d{\bf b},
    \label{afr_tripexp01int01}
\end{equation}
and since $1\big/{\left(1-z(1+i{\bf b\cdot B}) \right)^2}=(1-z)^{-2}\left[1+i\frac{z}{1-z}{\bf b\cdot B}\right]^2\big/\left[1+\frac{z^2}{(1-z)^2}({\bf b\cdot B})^2)\right]^2$
we obtain for Eq.~(\ref{afr_tripexp01int01}) (after making $\frac{z}{1-z}{\bf b}={\bf b'}$ substitution) 
\begin{equation}
\left(1-z\right)^{d-2}
\left(\frac{1}{2\pi}\right)^d \int_{-\frac{z\pi}{1-z}}^{\frac{z\pi}{1-z}}\int_{-\frac{z\pi}{1-z}}^{\frac{z\pi}{1-z}}\dots\int_{-\frac{z\pi}{1-z}}^{\frac{z\pi}{1-z}}
\frac{\left[1+i{\bf b'\cdot B}\right]^2}{\left[1+({\bf b'\cdot B})^2)\right]^2}d^d{\bf b'}.
    \label{afr_tripexp01int02}
\end{equation}
We see that in the $z\to 1$ limit the $z$ dependence arrives from the $(1-z)^{d-2}$ pre-factor and the fact that the range of integration diverges as $1/(1-z)$. For $d=1$ extra caution is needed since the pre-factor $1/(1-z)$ diverges while the integral $\int_{-\infty}^{\infty}\left[1+ib'B\right]^2\big/\left[1+(b'B)^2\right]^2db'=0$. Exact calculation of the integral in Eq.~(\ref{afr_tripexp01int02}) for $d=1$ shows that
\begin{equation}
\frac{1}{2\pi(1-z)}\int_{-\frac{z\pi}{1-z}}^{\frac{z\pi}{1-z}} \frac{[1+ib'B]^2}{\left[1+(b'B)^2\right]^2}d\,b'=\frac{1}{1+z(z-2+B^2\pi^2 z)}\underset{z\to 1}{\longrightarrow}
\frac{1}{B^2\pi^2}
    \label{afr_tripexp01d1}
\end{equation}
a constant and is not diverging in the $z\to 1$ limit. This proofs that for $d=1$ and the case of a present bias ($B\neq 0$) the sum $\sum_{\bf x'}{\hat c}_z({\bf x'}){\hat c}_z(-{\bf x'})$ converges to a constant when $z\to 1$ so the double sum $\sum_{\bf x}\sum_{\bf x'}\dots$ diverges as $1/(1-z)$ (and not as $1/(1-z)^2$) in the $z\to 1$ limit.
For any $d\geq 2$ the pre-factor $(1-z)^{d-2}$ in Eq.(\ref{afr_tripexp01int02}) is not diverging and the only divergences are possible from the range of the integration when $z\to 1$. 
Inspection of the function $[1+i \sum_{j=1}^d b'_j B_j]^2\big/\left[1+(\sum_{j=1}^d b'_j B_j)^2\right]^2$ shows that when the $|b'_1|\to\infty$ the leading order of this function is $\sim 1/(b'_1B_1+\sum_{j=2}^db'_j B_j)^2$.
Integration over $b'_1$ provides a leading order of $1/(b'_2 B_2+\sum_{j=3}^db'_j B_j)^1$ for $|b'_2|\to\infty$. 
Next integration over $b'_2$ will provide a leading order of $\log\left(\sum_{j=3}^d b'_j B_j\right)$ for the other $b'_j$s. 
By continuing the integration over all the different $b'_j$ ($d$ integrals in total) we obtain that the integrals in Eq.~(\ref{afr_tripexp01int02}) are diverging as $|(1-z)^{2-d}\log\left(1-z\right)|$ when $z\to 1$.
Then from Eq.~(\ref{afr_tripexp01d1}), Eq.~(\ref{afr_tripexp01int02}) and Eq.~(\ref{afr_triple01}) it is established that
\begin{equation}
\sum_{\bf x'}{\hat c}_z({\bf x'}){\hat c}_z(-{\bf x'})
\underset{z\to 1}{\sim} 
\left\{
\begin{array}{ll}
 \frac{1}{B^2\pi^2}    & d=1  \\
  |\log\left(1-z\right)|   &  d\geq2
\end{array}
\right.
    \label{afr_triple01fin}
\end{equation}
Finally we have shown that for any dimension of the lattice $d$, when the RW has a bias (i.e. ${\bf B\neq 0}$), the double sum 
$\sum_{\bf x}\sum_{\bf x'} {\hat c}_z({\bf x})
{\hat c}_z({\bf x'-x}){\hat c}_z({\bf x-x'})$ is growing as $|\log\left(1-z\right)|/(1-z)$ in the $z\to 1$ limit. 

The second type of behavior is the case without bias, i.e. 
\begin{equation}
{\hat p}({\bf b})\sim 1-\left({\bf b \cdot B}\right)^2 \qquad {\bf b}\to 0.
    \label{afr_tripexp02}
\end{equation}
In a similar fashion as Eq.~(\ref{afr_tripexp01int02}) was derived, we obtain that
\begin{equation}
\sum_{\bf x'}{\hat c}_z({\bf x'}){\hat c}_z(-{\bf x'})
\underset{z \to 1}{\longrightarrow} 
\left(1-z\right)^{d/2-2}
\left(\frac{1}{2\pi}\right)^d \int_{-\sqrt{\frac{z}{1-z}}\pi}^{\sqrt{\frac{z}{1-z}}\pi}\dots\int_{-\sqrt{\frac{z}{1-z}}\pi}^{\sqrt{\frac{z}{1-z}}\pi}
\frac{1}{\left[1+({\bf b'\cdot B})^2)\right]^2}d^d{\bf b'}.
    \label{afr_tripexp02int01}
\end{equation}
The integral on the right hand side of Eq.~(\ref{afr_tripexp02int01}) is always positive and the integration coordinates can be transformed into generalized polar coordinates. In this case the only non-constant integration is of the form $\int_0^{\sqrt{\frac{z}{1-z}}\pi|{\bf B}|}r^{d-1}/(1+r^2)^2$ that is diverges as $(1-z)^{2-d/2}|\log(1-z)|$ for $d\geq4$ and converges for any $d<4$. Eventually in the $z\to 1$ limit
\begin{equation}
    \sum_{\bf x'}{\hat c}_z({\bf x'}){\hat c}_z(-{\bf x'})
\underset{z \to 1}{\sim} \left\{
\begin{array}{ll}
(1-z)^{-3/2} & d=1 \\
(1-z)^{-1}  & d=2   \\
|\log\left((1-z)\right)| & d>2
\end{array}
\right.
    \label{afr_trip02fin}
\end{equation}
We have shown that for any dimension $d>2$, when the RW has no bias (i.e. ${\bf B= 0}$), the double sum 
$\sum_{\bf x}\sum_{\bf x'} {\hat c}_z({\bf x})
{\hat c}_z({\bf x'-x}){\hat c}_z({\bf x-x'})$ is growing as $|\log\left(1-z\right)|/(1-z)$ in the $z\to 1$ limit.

We have proven that for the specific case of $\sum_{\bf x}\sum_{\bf x'} {\hat c}_z({\bf x})
{\hat c}_z({\bf x'-x}){\hat c}_z({\bf x-x'})$ and transient RW the double sum diverges slower than $(1-z)^2$ in the $z\to 1$ limit. This result holds also for any double summation over ${\bf x}$ and ${\bf x'}$ and triple multiplications of the probability densities ${\hat c}_z({\bf x-x'}){\hat c}_z({\bf x'-x}){\hat c}_z({\bf x'})$ (or any permutation of the positions). Again, due to the properties of the convolution integrals that lead to Eqs.(\ref{afr_triple01fin},\ref{afr_trip02fin}). When the double summation is performed over multiplication of more than three ${\hat c}_z({\bf x})$s the result will be equivalent to several convolutions integral. Since each convolution reduces the order of divergence of $1/(1-z)$, additional convolutions will only reduce the divergences that appear in Eqs.~(\ref{afr_triple01fin},\ref{afr_trip02fin}). This means that the results of this section show that {\em any double summation over ${\bf x}$ and ${\bf x'}$ and n-th multiplication of positional PDFs diverges slower than $1/(1-z)^2$ when $z\to 1$, if the RW is transient}.

\bibliographystyle{apsrev4-1} 
\bibliography{./quenchedLiterature} 

\end{document}